\documentclass{aa}  

\usepackage{graphicx}
\usepackage{txfonts}
\newcommand{\emy}{\epsilon_\nu(250 \,\mu\mathrm{m})}
\newcommand{\emyrat}{I_\nu(250 \,\mu\mathrm{m})/I_\nu(500 \,\mu\mathrm{m})}

\begin{document} 

\title{Dust emissivity in resolved spiral galaxies\thanks{This work makes use of the DustPedia database. DustPedia is a collaborative focused research project supported by the European Union under the Seventh Framework Programme (2007- 2013) call (proposal no. 606824, P.I.\ J. I. Davies). The database is publicly available at {\tt http://dustpedia.astro.noa.gr}.} }


   \author{
Simone Bianchi\inst{1}
\and
Viviana Casasola\inst{2}
\and
Edvige Corbelli\inst{1}
\and
Fr\'ed\'eric Galliano\inst{3}
\and
Laura Magrini\inst{1}
\and
\\
Angelos Nersesian\inst{4,5}
\and
Francesco Salvestrini\inst{1}
\and 
Maarten Baes\inst{5}
\and
Letizia P.~Cassar\`a\inst{6}
\and
Christopher J. R.~Clark\inst{7}
\and
\\
Ilse De Looze\inst{5}
\and
Anthony P.~Jones\inst{8}
\and 
Suzanne C.~Madden\inst{3}
\and
Aleksandr Mosenkov\inst{9}
\and
Nathalie Ysard\inst{8}
}

\institute{
INAF - Osservatorio Astrofisico di Arcetri, Largo E. Fermi 5, I-50125, Florence, Italy\\
              \email{simone.bianchi@inaf.it}
\and
INAF - Istituto di Radioastronomia, Via P. Gobetti 101, I-40129, Bologna, Italy
\and
AIM, CEA, CNRS, Universit\'e Paris-Saclay, Universit\'e Paris Diderot, Sorbonne Paris Cit\'e, F-91191 Gif-sur-Yvette, France
\and
National Observatory of Athens, Institute for Astronomy, Astrophysics, Space Applications and Remote Sensing,  Ioannou Metaxa and Vasileos Pavlou GR-15236, Athens, Greece
\and
Sterrenkundig Observatorium, Universiteit Gent, Krijgslaan 281 S9, B-9000 Gent, Belgium
\and
INAF - Istituto di Astrofisica Spaziale e Fisica Cosmica, Via Alfonso Corti 12, 20133 Milan, Italy 
\and
Space Telescope Science Institute, 3700 San Martin Drive, Baltimore, Maryland 21218-2463, United States of America
\and
Universit\'e  Paris-Saclay, CNRS, Institut d'Astrophysique Spatiale, 91405 Orsay, France
\and
Department of Physics and Astronomy, N283 ESC, Brigham Young University, Provo, UT 84602, USA
}

   \date{}

 
  \abstract
{The far-infrared (FIR) and sub-millimeter (submm) emissivity, $\epsilon_\nu$, of the Milky Way (MW) cirrus is an important benchmark for dust grain models. Dust masses in other galaxies are generally derived from the FIR/submm using the emission properties of these MW-calibrated models.}
   {
We seek to derive the FIR/submm $\epsilon_\nu$ in nine nearby spiral galaxies to check its compatibility with MW cirrus measurements. 
     }
   {
We obtained values of $\epsilon_\nu$ at 70 to 500 $\mu$m, using maps of dust emission from the {\em Herschel} satellite and of gas surface density from the THINGS and HERACLES surveys on a  scale generally corresponding to 440 pc. We studied the variation of $\epsilon_\nu$ with 
the surface brightness ratio
$\emyrat$,  a proxy for the intensity of the interstellar radiation field heating the dust.
}
   {
We find that  the average value of $\epsilon_\nu$ agrees with MW estimates for pixels sharing the same color as the cirrus, namely, for $\emyrat = 4.5$. For $\emyrat > 5,$ the measured 
emissivity is instead up to a factor $\sim$2 lower than predicted from MW dust models heated by stronger radiation fields. Regions with higher $\emyrat$ are preferentially
closer to the galactic center and have a higher overall (stellar+gas) surface density and molecular fraction. The results do not depend strongly on the adopted CO-to-molecular 
conversion factor and  do not appear to be affected by the mixing of heating conditions.  }
 {
Our results confirm the validity of MW dust models at low density, but are at odds with predictions for grain evolution in higher 
density environments. If the lower-than-expected $\epsilon_\nu$ at high $\emyrat$ is the result of intrinsic variations in the dust 
properties,  it would imply an underestimation of the dust mass surface density of up to a factor $\sim$2 when using current dust models.
}

              
\titlerunning{Dust emissivity in resolved spiral galaxies}
   \maketitle
%

\section{Introduction}

\begin{table*}
\caption{Properties of the sample.}              
\label{tab:sample}      
\small
\centering                                      
\begin{tabular}{l c c c c c c c c }          
\hline\hline                        
name &  Hubble stage  & distance & $R_{25}$ &       scale&contribution &inclination&$12+\log(\mathrm{O/H})$            \\
          &  T                     & Mpc        &  arcmin    &   pc/pixel&to the sample$^a$ &degrees    & at 0.4$\times R_{25}$ \\
\hline                                   
NGC~628  (M74)           &5.2& 10.1 & 5.0      & 590 &         0.10&31& 8.56 \\
NGC~925                      &7.0&   9.2 & 5.4      & 540 &         0.06&67 & 8.29\\
NGC~2403                    &6.0&   3.2 &  10      & 190 &         0.08&64& 8.29\\
NGC~3521                    &4.0& 12.4 & 4.2      & 720 &         0.05&64& 8.63\\
NGC~4736 (M94)          &2.3&  4.4 &  3.9      & 260 &         0.05&34& 8.55\\
NGC~5055 (M63)          &4.0&  9.0 &  5.9      & 520 &         0.07&57& 8.69\\
NGC~5194 (M51)$^{b}$&4.0&  8.6 &  6.9      & 500 &         0.09&33& 8.72\\
NGC~5457 (M101)        &5.9&  7.1 &  12       & 410 &         0.37&22& 8.43\\
NGC~6946                    &5.9&  6.7 &  5.7      &  390 &        0.13&35& 8.59\\
\hline                                             
\end{tabular}
\tablefoot{
Hubble stage, distance, B-band optical radius $R_{25}$, and inclination are from the tabulated values of \citet{ClarkA&A2018} and \citet{MosenkovA&A2019}, available from
the DustPedia database. Characteristic metallicity at $0.4\times R_{25}$ have been derived from \citet{PilyuginAJ2014}.
\\
\tablefoottext{a}{ratio between the number of pixels analyzed for each galaxy and the total number of pixels in the sample.}
\\
\tablefoottext{b}{We excluded the companion NGC~5195 by cutting all pixels with $\delta \ge 47.24^\circ$.}
}
\end{table*}

Since the 1980s, a variety of satellites have carried out detailed studies of the thermal emission of dust grains in the Milky Way (MW), from its  far-infrared (FIR) peak
down to the submillimeter (submm) and longer wavelengths. In particular, observations of dust have been compared with those of hydrogen emission lines in order to derive 
the dust emissivity, $\epsilon_\nu$, that is, the surface brightness per H column density (see, e.g., \citealt{BoulangerA&A1996,DwekApJ1997,PlanckIntermediateXVII}). Besides the optical properties, size distribution, material 
composition, grain shape, and abundance, $\epsilon_\nu$ depends on the intensity of the interstellar radiation field (ISRF) heating the dust grains. For clouds at
high 
Galactic latitude, that is, the so-called cirrus \citep{LowApJL1984},  the starlight intensity is well known from observations and the local ISRF \citep[LISRF;][]{MathisA&A1983} can be used to estimate
the temperatures attained by grains of different materials and radii. Thus, the MW cirrus emissivity has become a benchmark in defining
the absorption (emission) cross-section of the dust grain mixtures (see \citealt{HensleyApJ2021} for a recent review on $\epsilon_\nu$ and other 
observational constraints for dust models).

Models of the MW dust, such as those of  \citet{ZubkoApJS2004},  \citet{DraineApJ2007b}, and The Heterogeneous dust Evolution Model for Interstellar Solids 
\citep[THEMIS;][]{JonesA&A2013,JonesA&A2017} are routinely applied to FIR-submm observations of other galaxies in order to extract various information. Such data include, for example, the amount of dust mass and the heating contribution from young stars, providing clues on the evolution of the metals constituting
the grains and extinction-free estimates of the star formation rate. Several works use MW dust models to fit the dust emission 
spectral energy distribution (SED) in galaxies, among them \citet{HuntA&A2019}, \citet{NersesianA&A2019}, \citet{AnianoApJ2020}, and
\citet{GallianoA&A2021}. The works cited above make use of data from the {\em Herschel} Space Observatory 
\citep{PilbrattA&A2010}, which has considerably improved the knowledge of the FIR-to-submm SED of galaxies, thanks to the mapping 
capabilities of the  Photodetector  Array Camera and Spectrometer \citep[PACS;][]{PoglitschA&A2010}, and the Spectral and Photometric 
Imaging Receiver \citep[SPIRE;][]{GriffinA&A2010}.

The question that remains is whether MW dust is representative  of the average grain properties in other galaxies. A first
test was attempted by \citet{JamesMNRAS2002}, who derived the disk-averaged dust absorption cross-section in nearby spirals, 
using estimates of the mass available in metals to infer the dust mass independently from FIR-submm observations. 
In a similar fashion,
\citet{ClarkMNRAS2016} and \citet{BianchiA&A2019} used {\em Herschel} global photometry and found results that are generally consistent with the MW properties, with a large scatter, however. Applying the same method pixel-by-pixel to {\em Herschel} maps of
two galaxies, \citet{ClarkMNRAS2019} reported variations of the absorption cross-section with the environment, with smaller absorption
cross-sections for pixels characterized by larger interstellar medium (ISM) densities, apparently at odds with models of the evolution of
dust in denser environments \citep[e.g.,][]{KoehlerA&A2015}.

However, the process of estimating the dust mass (or surface density in resolved analysis) from metals, following the method of 
 \citet{JamesMNRAS2002},  is subject to several uncertainties: i) the total gas column 
density is needed, which relies on the recipe used to derive its molecular component, typically from observations of CO lines; ii) an estimate of the 
dust-to-gas mass ratio must be inferred from metallicity measurements, from solar or local ISM values,
taking into account the observed metal depletions; iii) a functional form needs to be assumed for the cross section  (a power law) and a single
temperature is estimated from fitting the SED, thus basically assuming that dust grains have the same properties everywhere in a galaxy. Indeed,
\citet{PriestleyMNRAS2020} have suggested that the result of \citet{ClarkMNRAS2019} could be biased by dust at colder temperatures,
which is expected in denser environments because there the ISRF is attenuated by dust extinction. This colder dust would be unaccounted 
for in a single-temperature fit.

To overcome these uncertainties, \citet{BianchiA&A2019} suggested limiting the analysis to the emissivity, $\epsilon_\nu$, for which no temperature
derivation and dust-to-gas prescription are needed; it is only the uncertainty in the derivation of the molecular gas from CO that remains. The emissivity $\epsilon_\nu$
can then be compared with the response of a dust model (obtained either from fitting the MW cirrus data or after following the evolution of grains in 
different environments) to ISRFs of different intensities. For their sample of 204 objects,  \citet{BianchiA&A2019} found an average emissivity 
similar to that measured in the MW, for the same FIR color as in the cirrus, and a reduction in emissivity (with respect to model predictions) for galaxies with 
more intense ISRF and a larger contribution of molecular gas to the ISM (thus, similarly to that found by \citealt{ClarkMNRAS2019},
given that higher molecular fractions tends to be closely associated with higher gas surface densities). 
The dependence on the interstellar environment could only be studied indirectly by \citet{BianchiA&A2019}, however, since a single global emissivity was derived for each object.

Improving on the work of \citet{BianchiA&A2019}, here we derive the emissivity using dust emission and gas column density maps for a sample of nine low-inclination spirals.
These objects are among the largest and nearest galaxies observed by {\em Herschel} and are thus included in the DustPedia sample  \citep{DaviesPASP2017}.
The aim of the analysis is to verify, on a resolved scale,  the general validity of the MW dust model for other galaxies and to assess possible variations of the dust properties
across galactic disks. 

The paper is organized as follows: in Sect.~\ref{sec:sample}, we describe the sample and dataset we use. The emissivity derivation is presented in 
Sect.~\ref{sec:method}. Results on $\emy$ are shown in Sect.~\ref{sec:emy} and discussed in Sect.~\ref{sec:discussion}. The emissivity SED is described in Sect.~\ref{sec:spectrum}. In Sect.~\ref{sec:summary}, we summarize the results and present our conclusions.

\section{Sample and dataset}
\label{sec:sample}

Our sample is drawn from the work of \citet{CasasolaA&A2017}, who selected the largest low-inclination spiral galaxies in DustPedia (18 objects more extended than 9$\arcmin$ in SPIRE images) and study the spatial distribution of stellar emission, dust emission, gas column density, and other derived quantities. 
Among these, we chose objects that have  both observations of atomic and molecular gas from homogeneous sources.
We retained 9 galaxies: they are listed in Table~\ref{tab:sample}.

{\em Herschel} dust  continuum images from PACS at 70, 100 and 160 $\mu$m and from SPIRE at 250, 350 and 500 $\mu$m were taken from the DustPedia database \citep{ClarkA&A2018}. Map resolutions range from FWHM$\approx 9\arcsec$ at 70  $\mu$m to $36\arcsec$ at 500 $\mu$m. The majority of galaxies have the full FIR/submm wavelength coverage, with the exceptions of  NGC~2403 and NGC~5194, which were not observed at 100 $\mu$m. 
As in \citet{CasasolaA&A2017}, we used sky areas outside the targets to estimate the instrumental and confusion noise.

Maps of the \ion{H}{i} column density were obtained from 21 cm line observations undertaken at  the Very Large Array of  the  National  Radio  Astronomy  Observatory, within The HI Nearby Galaxy Survey \citep[THINGS;][]{WalterAJ2008}. We used the ROBUST dataset, providing a uniform synthesized beam with FWHM $\approx 6\arcsec$. The original moment 0 maps (and their estimated noise) were converted into units of column density with the usual assumption of optically thin \ion{H}{i} emission.

To trace the molecular gas column density, we used the moment 0 maps (and error maps) from the HERA CO-Line Extragalactic Survey  \citep[HERACLES;][]{LeroyAJ2009}, which observed the
$^{12}$CO(2-1) line  at  the  IRAM  30  m  telescope with resolution FWHM=11$\arcsec$.

We smoothed all images to the lowest resolution of the dataset, namely, FWHM=36$\arcsec$ at 500 $\mu$m. We used the convolution kernels of \citet{AnianoPASP2011}: for {\em Herschel} maps, kernels are available to pass from the original point spread function (PSF) to that at 500 $\mu$m; for gas column density maps, we assumed that the original PSF is a Gaussian of appropriate size. All maps were regridded to a common grid of pixel size 12$\arcsec$ (the pixel scale of SPIRE 500 $\mu$m images in DustPedia, equivalent to FWHM/3).

As detailed in the next section, we applied metallicity and column density-dependent CO-to-H$_2$ conversion factors: for this, we used the O/H metallicity gradients provided by  \citet[][see Appendix~\ref{app:metgrads} for a justification]{PilyuginAJ2014} and the stellar surface density maps derived by \citet{CasasolaA&A2017}. For each galaxy, the total surface density was derived by adding together the stellar and gas contribution (with the usual 36\% correction to include He and other metals in the gas budget), and de-projecting the maps using the inclinations listed in Table~\ref{tab:sample}. We also used NUV maps from the GALaxy Evolution eXplorer (GALEX; \citealt{MorriseyApJS2007}) and 3.6 $\mu$m maps from the InfraRed Array Camera (IRAC) aboard the {\em Spitzer} Space Telescope \citep{WernerApJS2004},  available from the DustPedia database; these were corrected for foreground MW extinction as in \citet{ClarkA&A2018}.

\section{Method}
\label{sec:method}

For the MW cirrus, the dust emissivity $\epsilon_\nu$  is defined as the surface brightness of dust emission $I_\nu$ per hydrogen column density $N_\mathrm{H}$. For each pixel considered in our analysis, we estimated the emissivity accordingly, using
\begin{equation}
\epsilon_\nu = \frac{I_\nu}{N_\mathrm{H}} = \frac{I_\nu}{N_\mathrm{HI} + 2\times N_\mathrm{H_2}}.
\label{eq:emy}
\end{equation}
While the column density of atomic gas can be derived directly from observations of the 21 cm \ion{H}{i} line, $N_\mathrm{H_2}$ can only be estimated indirectly:
typically, the lines of CO are used, as they are considered the main tracer of molecular gas. Here, since we are using the intensity of the CO(2-1) line, we can write:
\begin{equation}
N_\mathrm{H_2} = X_\mathrm{CO} \times  \frac{ I_\mathrm{CO(2-1)} }{ R_{21}},
\label{eq:co_conv}
\end{equation}
where $R_{21} = I_\mathrm{CO(2-1)}/I_\mathrm{CO(1-0)} $ is the ratio between the intensities of the (2-1) and (1-0) lines, and $X_\mathrm{CO}$ is the CO-to-H$_2$ conversion factor, normally derived for the CO(1-0) line.

To maintain a consistency with \citet{BianchiA&A2019}, where molecular masses from \citet{CasasolaA&A2020} are used, here we adopt $R_{21}=0.7$, a typical value found in the MW disk and in some HERACLES galaxies  \citep{LeroyAJ2013}. This value is consistent with the recent determination of \citet{BrokMNRAS2021}, who found an average $R_{21}=0.63$ 
with a scatter of 15\%, on a sample of nine nearby galaxies (four of which in our sample), as well as with \citet{LeroyApJ2022}, who have  $R_{21}=0.65$ and a scatter of 30\% on a larger sample
of 43 objects (seven in common with ours). 
Both works find evidence of a  possible increase of $R_{21}$ in denser environments, and also find that the scatter is dominated by galaxy-to-galaxy variations. 

For the conversion factor, studies of molecular clouds in the MW disk have found:
\begin{equation}
X_\mathrm{CO}^\mathrm{MW} = 2.0 \times 10^{20} \mathrm{cm^{-2} \; (K\; km\; s^{-1})^{-1}},
\end{equation}
which is estimated to have an uncertainty of 30\%  \citep{BolattoARA&A2013}. The $X_\mathrm{CO}$ factor is supposed to increase in lower metallicity environments, because of the reduced availability of the CO molecule and because of reduced shielding against its photodissociation in less dusty clouds; instead, it should drop in higher density regions, such as galactic centers, where the gas temperature is higher.  Based on observations and theoretical models, \citet{BolattoARA&A2013} propose a metallicity and surface-density-dependent conversion factor, which can be written as:
\begin{equation}
X_\mathrm{CO}^\mathrm{B13} = \frac{X_\mathrm{CO}^\mathrm{MW}}{e^{0.4}} \times\exp\left(\frac{0.4}{Z/Z_\odot}\right) \times  \left( \frac{\Sigma_\mathrm{total}}{100 \; M_\odot \; \mathrm{pc}^{-2}}\right)^{-\gamma}, 
\label{eq:b13}
\end{equation}
with $\gamma=0.5$ for a total surface density (stars + gas) $\Sigma_\mathrm{total} > 100 \; M_\odot \; \mathrm{pc}^{-2}$ and $\gamma=0$ otherwise; 
the metallicity derived from (O/H) by using 
$\log_{10}(Z/Z_\odot)=12+\log_{10}(\mathrm{O/H})-8.69,$ so that $\log_{10}(Z/Z_\odot)=0$ for the MW disk environment (assuming it follows the Solar mixture). Equation~\ref{eq:b13} further assumes that the characteristic surface density of molecular clouds in a galaxy and in the MW is the same and equal to $100 \; M_\odot \; \mathrm{pc}^{-2}$ \citep[see also][]{ChiangApJ2021}.

Other works based on large galaxy samples suggest a power-law dependence of $X_\mathrm{CO}$ on global metallicity. In those studies, the power-law index is derived by imposing that the molecular mass must correlate with the star formation rate, for all metallicities. One such work is \citet{HuntA&A2020}, who found:
\begin{equation}
X_\mathrm{CO}^\mathrm{H20} = X_\mathrm{CO}^\mathrm{MW} \times
\left\{
    \begin{array}{ll}
\left( Z/Z_\odot \right)^{-1.55} & \mbox{if $Z<Z_\odot$},\\
      1 & \mbox{otherwise}.
    \end{array}
  \right.
\end{equation}
A similar power-law index was found by \citeauthor{AmorinA&A2016} (\citeyear{AmorinA&A2016}; this is used by  \citealt{BianchiA&A2019}), although steeper metallicity dependences have also been found \citep{HuntA&A2015,MaddenA&A2020}.

In this work, we assume the $X_\mathrm{CO}^\mathrm{B13}$ conversion factor. We also comment on the variation of the emissivity when $X_\mathrm{CO}^\mathrm{MW} $ and $X_\mathrm{CO}^\mathrm{H20}$ are used;
we warn, thought, that the latter was derived on global galactic properties and not on a resolved scale. 
 
We did not include the contribution of \ion{H}{ii} to $N_\mathrm{H}$: this is normally neglected in the derivation of the MW emissivity
and it is not included in the estimate of the average emissivities of DustPedia galaxies also  \citep[for a discussion, see][]{BianchiA&A2019}.

\section{Results}
\label{sec:emy} 

\begin{figure*}
\begin{center}
\includegraphics[width=17.3cm]{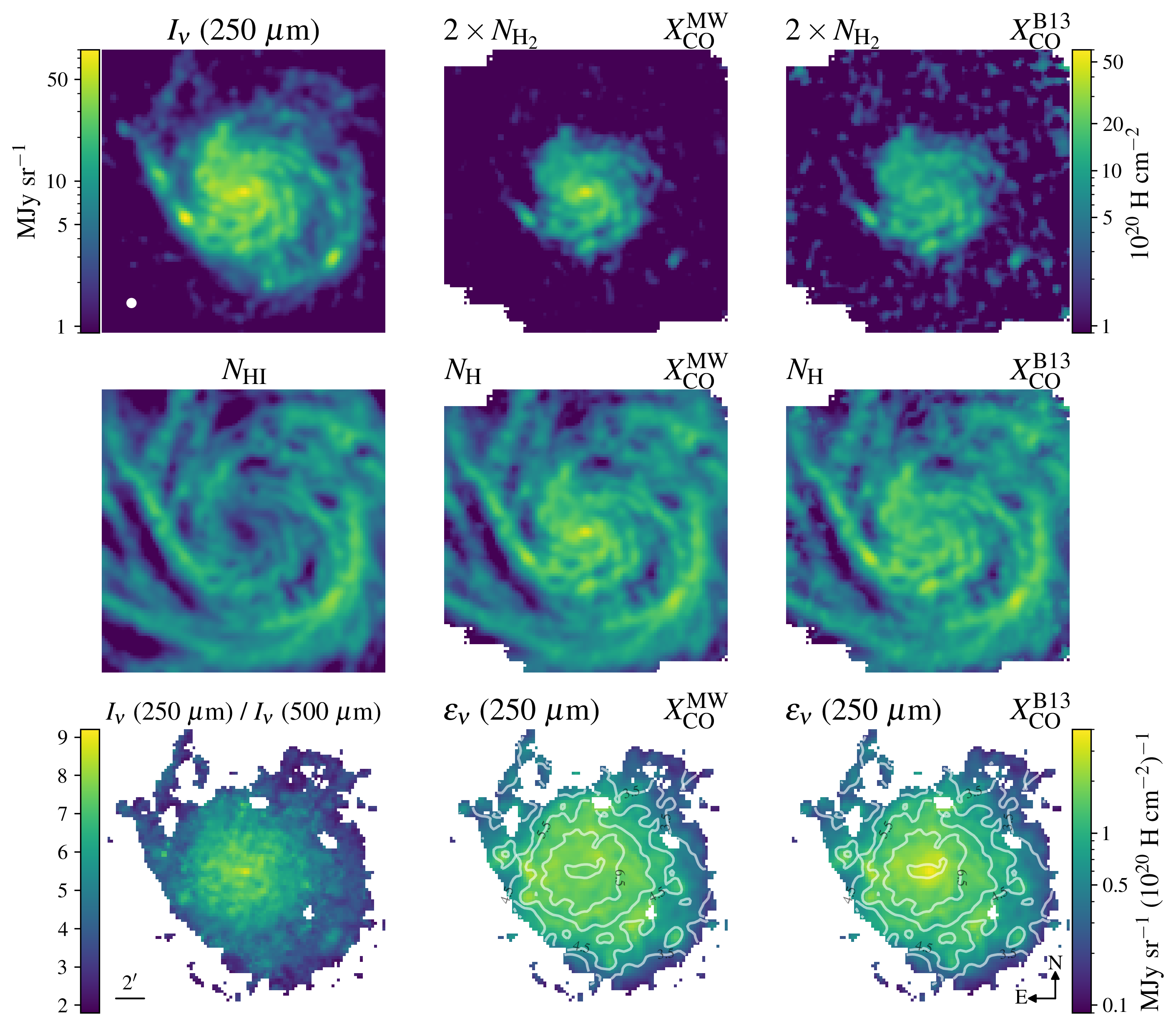}
\end{center}
\caption{
Dataset and results for NGC5457. The top row shows the 250 $\mu$m image, the molecular gas column density derived from CO observations for $X_\mathrm{CO}^\mathrm{MW}$, the same for $X_\mathrm{CO}^\mathrm{B13}$; the central row shows the atomic gas column density $N_\mathrm{\ion{H}{i}}$, and the total column density 
$N_\mathrm{H}$ for the two choices of $X_\mathrm{CO}$; the bottom row $\emyrat$, and $\emy$ for the two $X_\mathrm{CO}$. Maps of gas column densities have the same scale (top-right color bar), as well as the two $\emy$ maps (bottom-right color bar). Contours on the  $\emy$ maps show the central value of the $\emyrat$ bins used in the analysis. All images are convolved to FWHM $\approx36"$ (white circle in the top-left panel).
}
\label{fig:images}
\end{figure*}

\begin{figure*}
\begin{center}
\includegraphics[width=17.3cm]{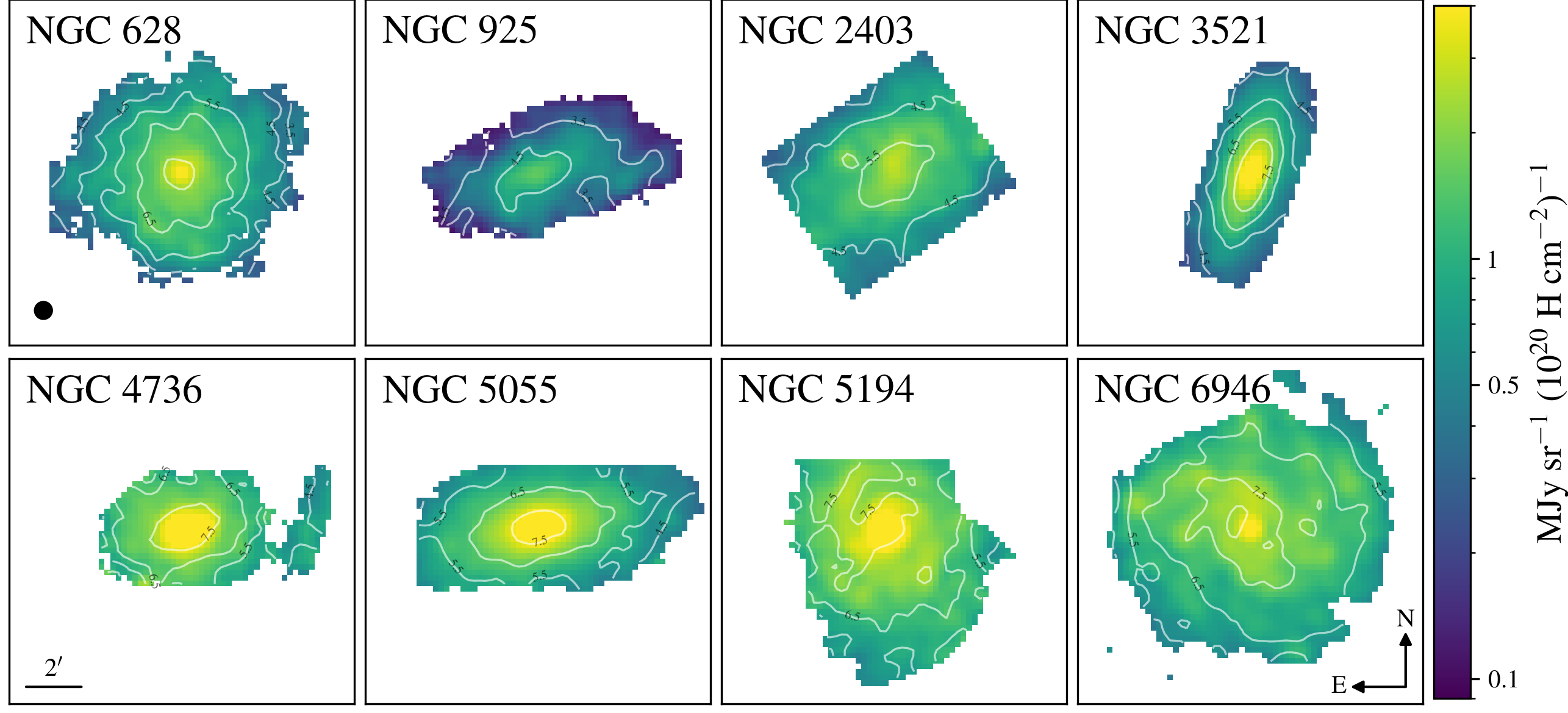}
\end{center}
\caption{
Maps of $\emy$ for the rest of the sample (assuming $X_\mathrm{CO}^\mathrm{B13}$).  Contours and FWHM as in Fig.~\ref{fig:images} (however, we note the different spatial scale).
}
\label{fig:otherimages}
\end{figure*}

For each galaxy in the sample, we derived the emissivity in all {\em Herschel} bands, using Eq.~\ref{eq:emy}. Following  \citet{BianchiA&A2019}, we conduct our analysis as a function of 
$\emyrat$, which, for standard models of dust emission, is directly related to the intensity of the starlight heating the dust grains (see Appendix~\ref{app:bias}).
Given Eq.~\ref{eq:emy}, $\emyrat$ is equivalent to $\epsilon(250 \,\mu\mathrm{m}) / \epsilon(500 \,\mu\mathrm{m})$. 
We used all pixels with $S/N > 3$ in the $\emyrat$ maps, defining a contiguous large area for each galaxy and allowing for the selection of high-quality data for all the other quantities we analyze here. 
In total, the whole sample consists of about 13,600 pixels. 
The contribution of each galaxy to the available pixels 
depends on its physical dimension and distance, but also on the limited coverage of the CO data for some objects. 
In Fig.~\ref{fig:images} we show the dataset, $\emy$, and $\emyrat$ for NGC~5457, the largest galaxy we analyze, contributing more than one-third of the available pixels  (see Table~\ref{tab:sample}).
Maps for $\emy$ with contours for  $\emyrat$  for the other galaxies of the sample are presented in Fig.~\ref{fig:otherimages}.

\begin{figure*}
\begin{center}
\includegraphics[scale=0.3]{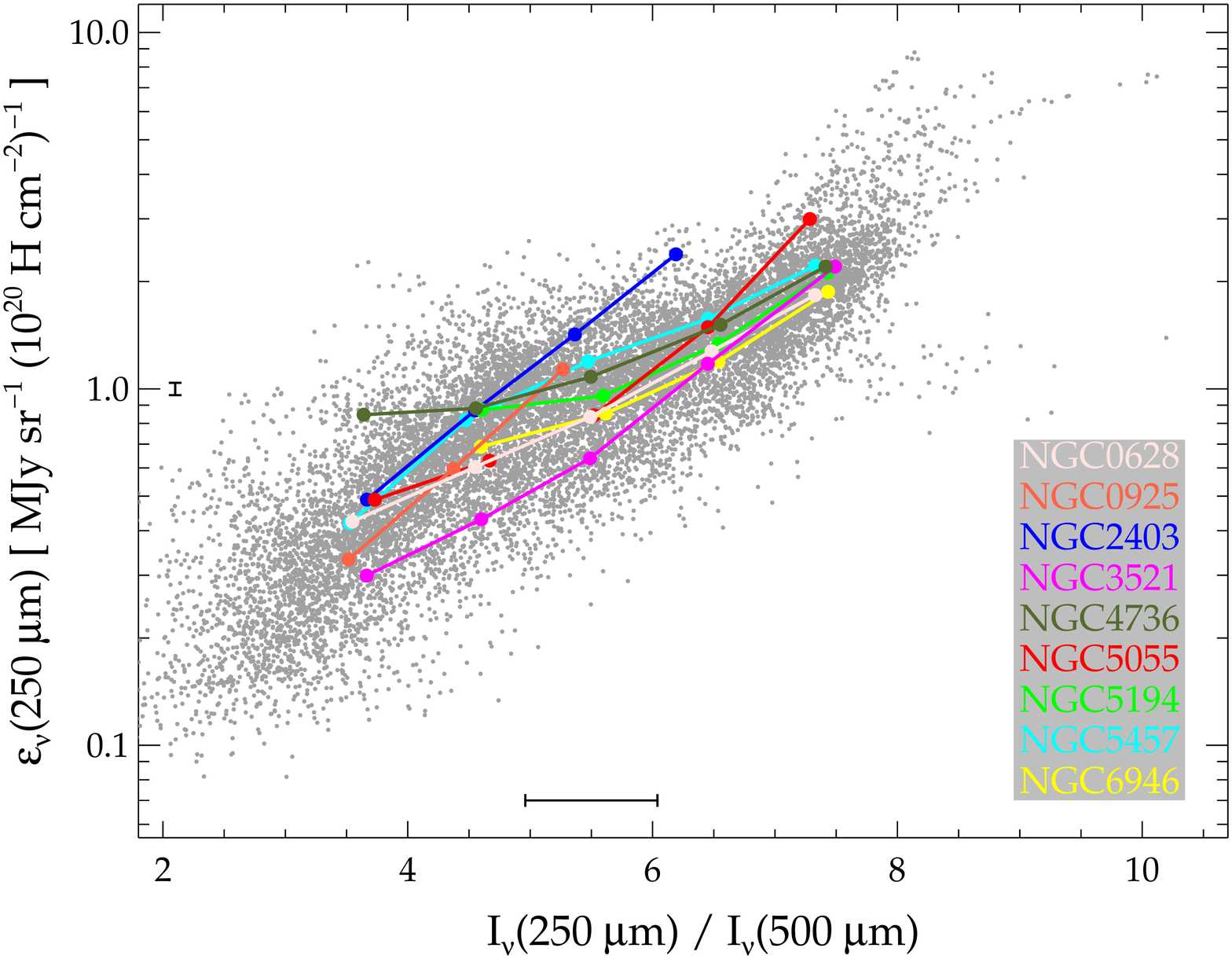}\includegraphics[scale=0.3, trim=105 0 0 0, clip]{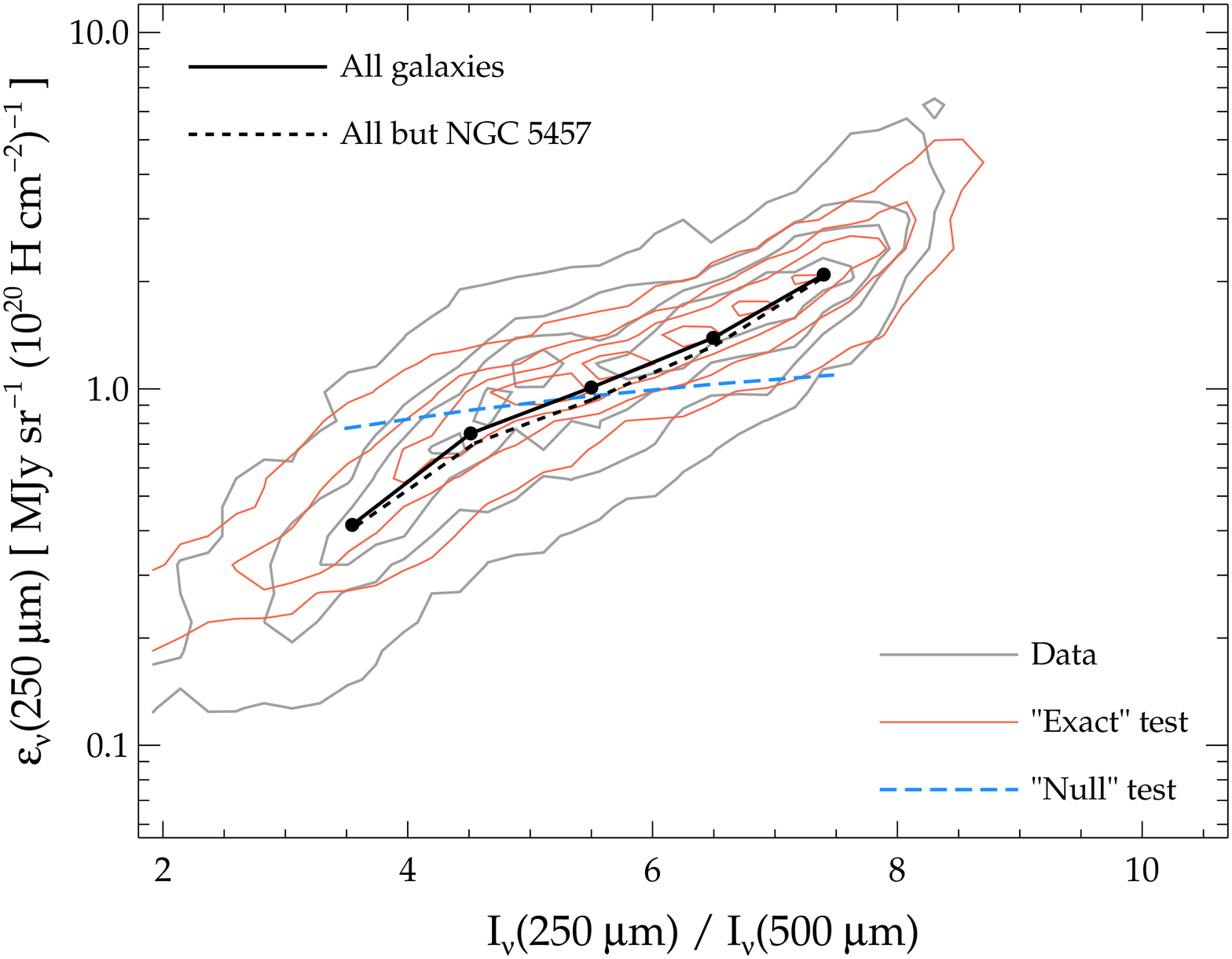}
\end{center}
\caption{
Emissivity at 250 $\mu$m, $\emy$ versus the $\emyrat$ ratio (assuming $X_\mathrm{CO}^\mathrm{B13}$). In the left panel, individual pixels are shown; in the right panel, gray contours encompass 95, 75, 50, and 25\% of the sample. The mean $\emy$ obtained after averaging pixels over five bins in the range of $3 \le \emyrat < 8$
is also shown: in the left panel, for each galaxy separately; in the right panel, for the total sample (and the total sample but NGC~5457). 
The left panel further shows the average errorbars for the central $\emyrat$ bin;
the right panel, the expectation for the "exact" (red contours) and "null" (blue binned averages) hypotheses.
}
\label{fig:emy250_xcob}
\end{figure*}

In Fig.~\ref{fig:emy250_xcob}, we show $\emy$ as a function of $\emyrat$, for all the pixels (left panel) and as contours of the pixel density (right panel).
As expected, $\emy$ grows with  $\emyrat$, as also found by \citet{BianchiA&A2019} for galaxy integrated values. 
In the left panel, we plot the mean value of $\emy$ 
for each galaxy, after binning the pixels in five equal bins of $\emyrat$ in the range: [3,8]. Individual objects have their own trend. In addition,  not all galaxies cover the full set of bins: NGC~5457 and NGC~628 have pixels spanning the whole $\emyrat$ range; in NGC~925 and NGC~2403 most of the pixels falls in the first two color bins; the rest of the galaxies, instead, privileges the higher  $\emyrat$ values (see also the contour in Fig.~\ref{fig:otherimages}).

The average $\emy$ of the whole sample is shown by the solid black line in the right panel of Fig.~\ref{fig:emy250_xcob}:
as in \citet{BianchiA&A2019}, we  used the mean of all pixels within each bin (without any weighting). 
The trend does not appear to be influenced strongly by a particular behavior of an individual galaxy: for example, it does not change significantly when we exclude the dominant contribution from NGC~5457 (dashed black line).
We also checked the impact of the different spatial scales of our galaxies: they range from 190 pc, for NGC~2403, to 720 pc, for NGC~3521 (for the adopted angular scale of 12$\arcsec$; Table~\ref{tab:sample}). The average spatial scale over the whole pixel sample is 440 pc. Despite the wide range, a large fraction of the available pixels ($\approx 70\%$) belongs to five galaxies with scales within 25\% of the average; the other four galaxies are the two nearest and two farthest ones.
If we restrict the analysis to the five galaxies with more uniform spatial scales, the average $\emy$ trend is almost  indistinguishable from that of the whole sample: thus,
440~pc can be considered the characteristic physical scale of our analysis.
Because of this, we have preferred to keep a fixed angular scale grid (as also done in other works; see, e.g., \citealt{SandstromApJ2013,JiaoMNRAS2021}), rather than to degrade all the scales to that of farthest object and significantly reduce the number of pixels.

The scatter around the mean values in Fig.~\ref{fig:emy250_xcob} is reduced with respect to that in \citet{BianchiA&A2019}: for the bin centered at  $\emyrat=4.5,$ it is $\approx 40$\% and $\approx 30$\% for  
$\emyrat=7.5$. Instead, it was larger for the integrated analysis, namely $\approx 50$\% in all bins.
The reduction in scatter is likely due to the more homogeneous dataset in the current work. \citet{BianchiA&A2019} rely on gas observations
gathered from several sources in the literature, with  gas masses (and thus average surface densities) obtained after applying general recipes  for aperture corrections
(in particular  for H$_2$ masses, the majority of which were derived from observations of  the sole galactic centre; see \citealt{CasasolaA&A2020}). 

Part of the scatter in Fig.~\ref{fig:emy250_xcob} is due to photometric errors: for $\emyrat$, the average uncertainty is $\approx 0.5$ for the first three bins, and $\approx 0.4$ for the last two; it rises to $\approx 0.5-0.6$ when the calibration error is taken into account (using the fraction of it which is uncorrelated among the SPIRE bands, see Appendix~A in \citealt{GallianoA&A2021}). This justifies the choice of averaging over bins of width $\Delta\emyrat$=1.
The average error on $\emy$ is instead a few percent (plus an additional 6\% when the calibration error at 250 $\mu$m is included; \citealt{GallianoA&A2021}). 
The average $\emy$ and $\emyrat$ errors for the central bin are shown in the left panel of Fig.~\ref{fig:emy250_xcob}.
In order to test the effects of these uncertainties on the pixel distribution in 
the $\emy$-$\emyrat$ plane, we made a simulation assuming that the average trend in Fig.~\ref{fig:emy250_xcob} is the "exact" behavior of emissivity in all galaxies. 
For each pixel, we used the measured $I_\nu(250 \,\mu\mathrm{m})$ to derive a mock $I_\nu(500 \,\mu\mathrm{m})$ according to the "exact" trend. From both flux densities we then extracted random values, adopting Gaussian errors and the measured uncertainty for that pixel (but neglecting calibration errors). The same was done for $N_\mathrm{H}$ (neglecting the uncertainties in the CO-to-H$_2$ conversion factor, which will be discussed later). Finally, mock realizations of $\emy$ and $\emyrat$ according to this "exact" hypothesis were produced: one of them is shown with red contours  in the right panel of Fig.~\ref{fig:emy250_xcob}. From this test, we obtain the result that 
photometric uncertainties alone can produce a scatter in $\emy$ of about 20\%, because of the shuffling due to the large bin (and uncertainties) in $\emyrat$, and of 
 the $\emy$ uncertainties themselves. After subtracting this contribution, we conclude that there is an intrinsic scatter in $\emy$ of $\approx$ 35\% and 20\% (at $\emyrat=4.5$ and 7.5, respectively) due to pixel-by-pixel and object-by-object variations in the dust properties.
The fact that scatter is not only due to statistical errors is also indicated by the differences between the individual galaxy averages as well as between those averages and the whole sample average:
these are much larger than the standard error of the mean (not shown, being smaller than the symbol size used for the average trends in Fig.~\ref{fig:emy250_xcob}). 

The random procedure of the previous paragraph was also used to check to what extent the trend seen in Fig.~\ref{fig:emy250_xcob}  is induced by the use of the same quantity, 
$I_\nu(250 \,\mu\mathrm{m})$, in the derivation of both $\emy$ and $\emyrat$. In fact, when non-independent quantities are plotted, spurious correlations might arise because of the uncertainties (see, e.g., \citealt{MosenkovMNRAS2014,DeVisMNRAS2017a}). We checked this by adopting a "null" hypothesis, namely, that there is no correlation between $I_\nu(250 \,\mu\mathrm{m})$ and $\emyrat$, and thus between  $I_\nu(250 \,\mu\mathrm{m})$ and $I_\nu(500 \,\mu\mathrm{m})$; this was done by shuffling the pixel values within the 500 $\mu$m maps. Indeed, we found that a spurious correlation can arise between  $\emy$ and $\emyrat$ and we show the average trend with a dashed blue line in the right panel of Fig.~\ref{fig:emy250_xcob}. However, this trend is much fainter than what is actually found in the data. Indeed, a physical connection between $\emyrat$ and $\emy$ is expected: 
in a warmer radiation field, dust is heated to higher temperatures and $\emyrat$ should increase; at the same time, under the thermal emission hypothesis, both $I_\nu(250 \,\mu\mathrm{m})$ and $I_\nu(500 \,\mu\mathrm{m})$ (and the related emissivities), should grow.

\begin{figure*}
\begin{center}
\includegraphics[scale=0.3, trim= 0 80 0 0,clip]{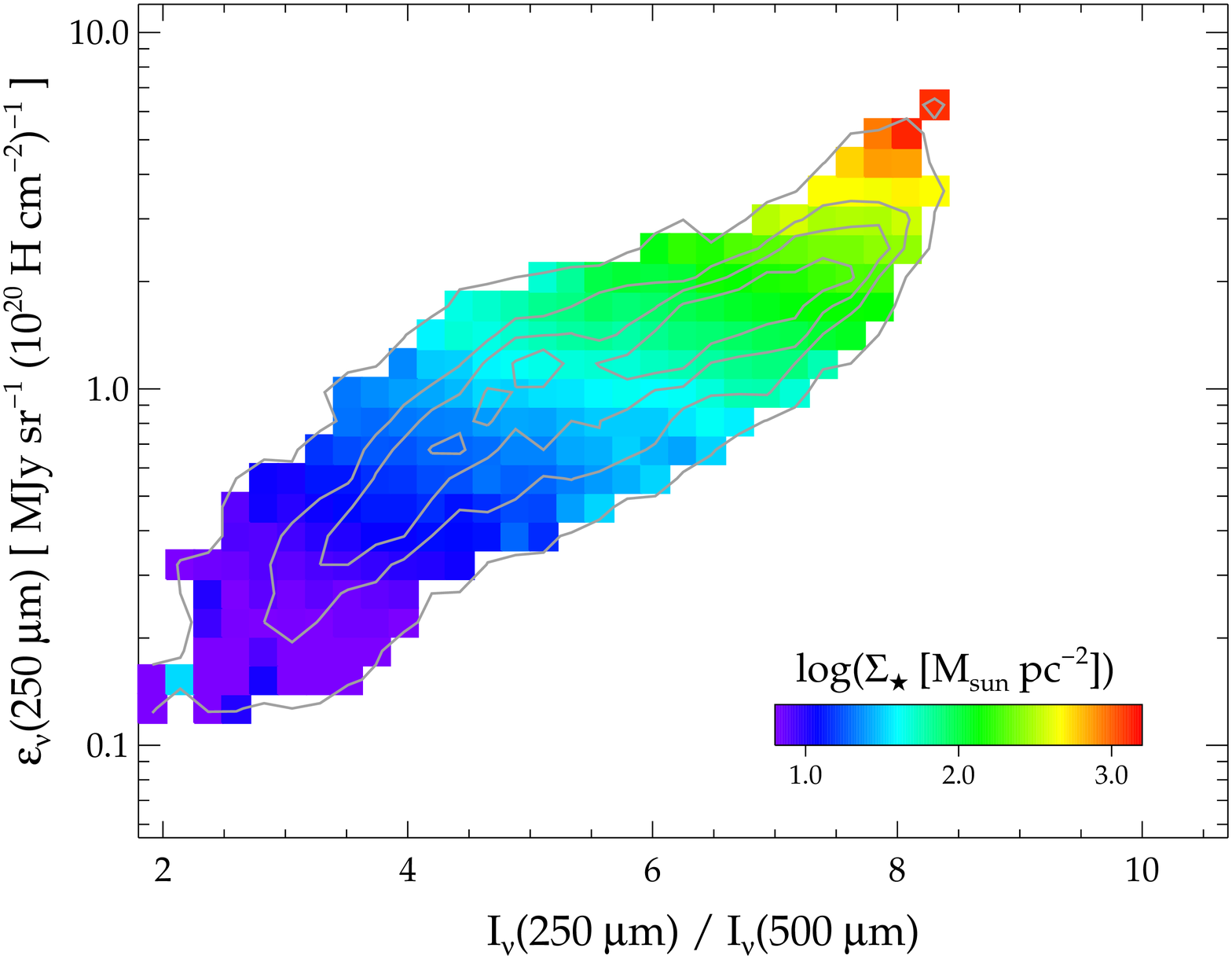}\includegraphics[scale=0.3,trim=105 80 0 0, clip]{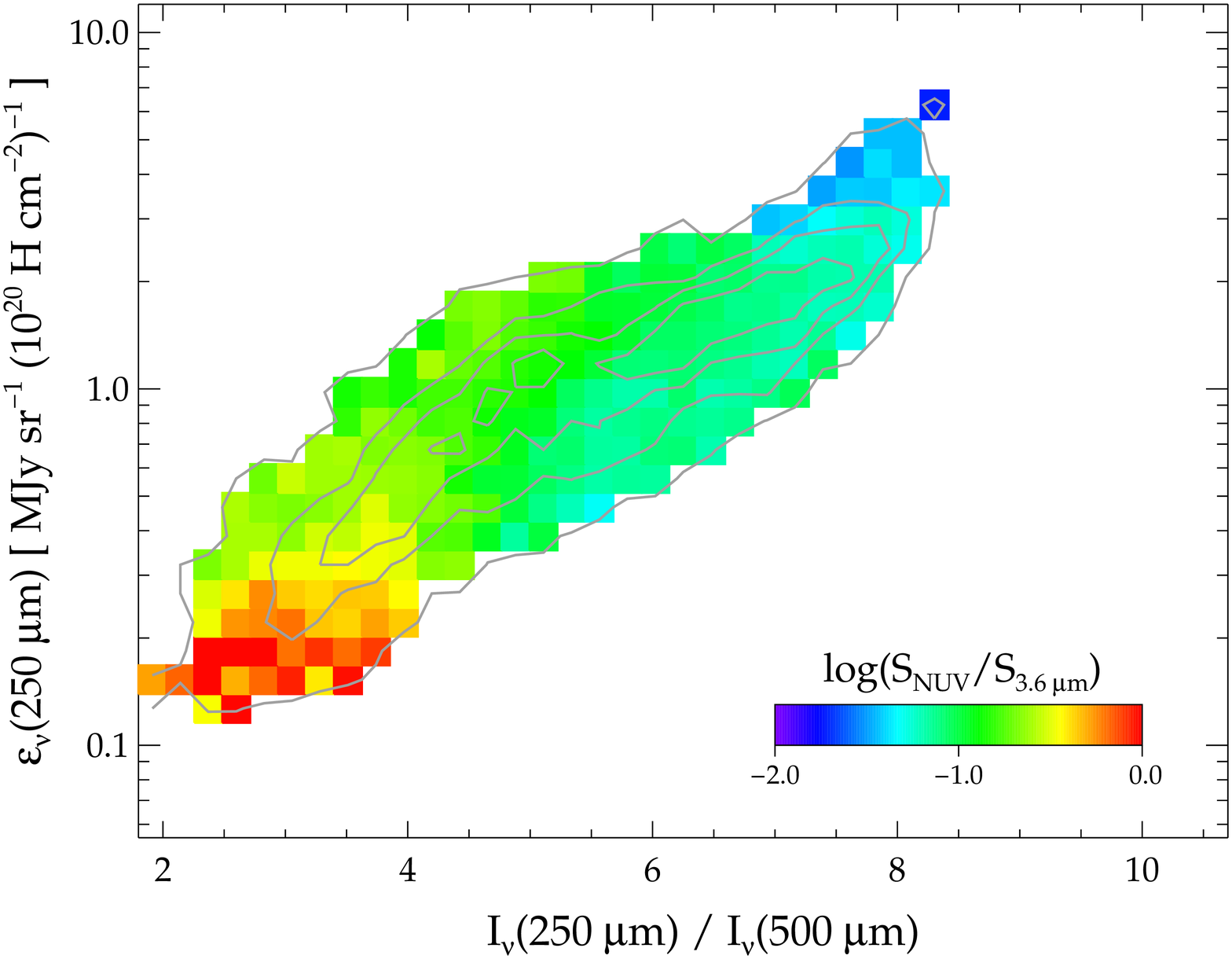}
\includegraphics[scale=0.3]{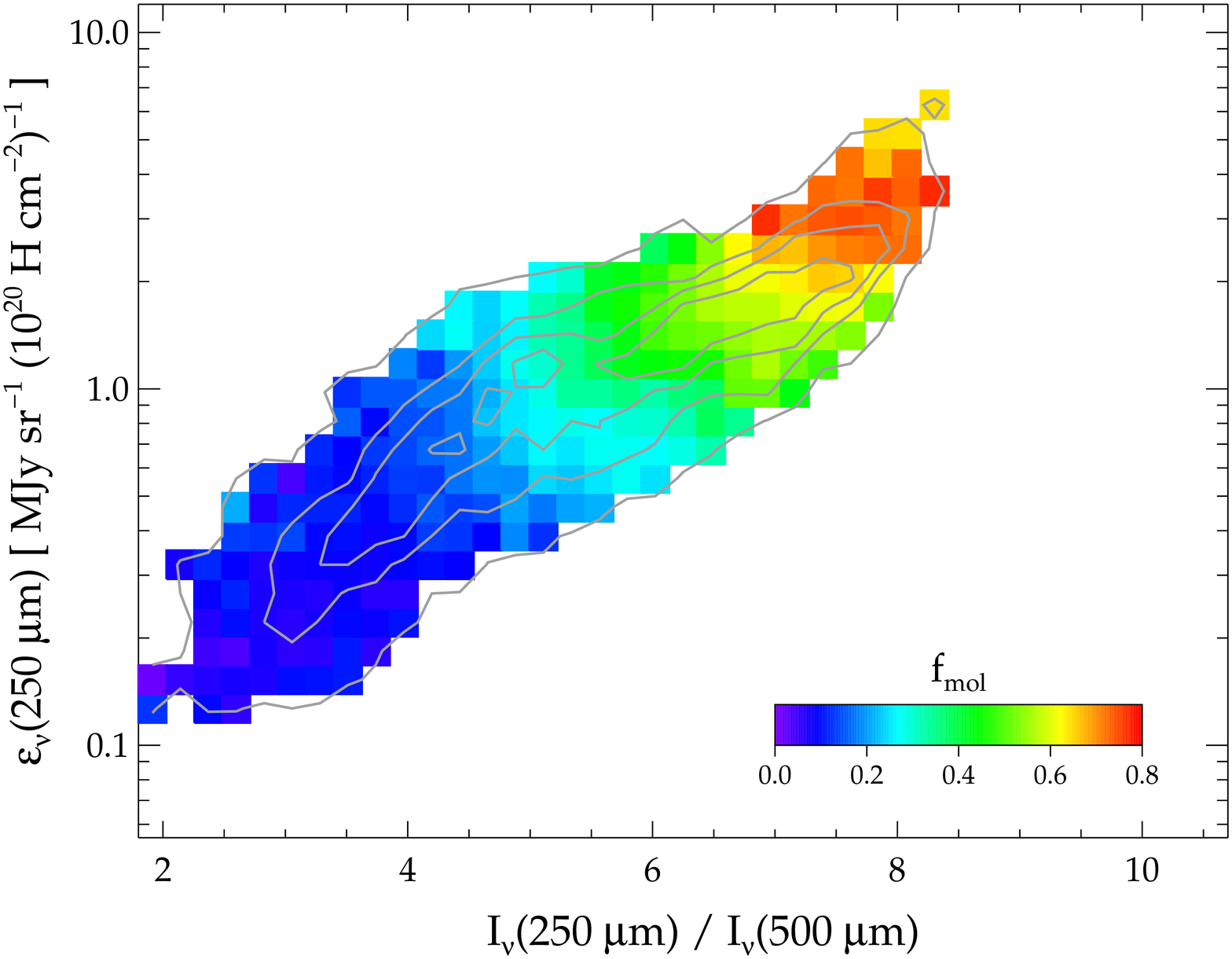}\includegraphics[scale=0.3, trim=105 0 0 0, clip]{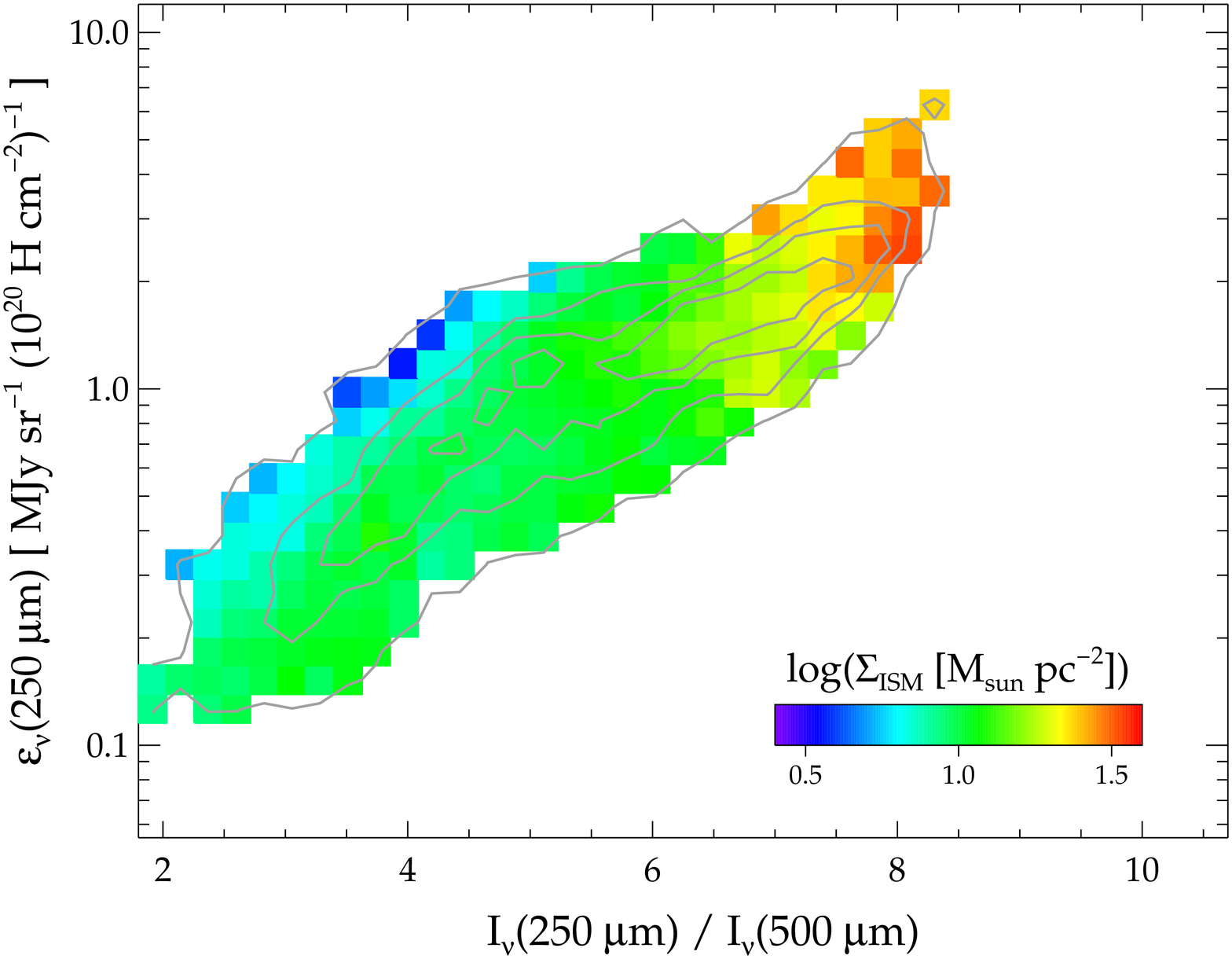}
\end{center}
\caption{
 $\emy$ versus $\emyrat$, same grey contours as in Fig.~\ref{fig:emy250_xcob}. In each panel, the color map shows the average value of the following quantities: $\Sigma_\star$, $S_\mathrm{NUV}/S_\mathrm{3.6\,\mu m}$,  $f_\mathrm{mol}$ and $\Sigma_\mathrm{ISM}$.
}
\label{fig:emy250_xcob_cmaps}
\label{fig:sism}
\end{figure*}

We also studied the dependence of the $\emy$-$\emyrat$ correlation on other quantities available through the DustPedia archive and previous analysis. For each galaxy, both $\emy$ and $\emyrat$ are higher in regions where the optical-NIR stellar emission, the gas column density, and the metallicity are higher.  However, we did not find any quantity that could be associated with a galaxy's individual trend in the  $\emy$-$\emyrat$ plane. Thus, in the following, we will only discuss the general trend for the whole of the sample. In  Fig.~\ref{fig:emy250_xcob_cmaps} we show, on the $\emy$-$\emyrat$ plane, the average values for the stellar mass surface density $\Sigma_\star$, the UV-to-NIR ratio $S_\mathrm{NUV}/S_\mathrm{3.6\,\mu m}$,  the fraction of molecular gas $f_\mathrm{mol}=2\times N_\mathrm{H_2}/N_\mathrm{H}$, and the gas mass surface density $\Sigma_\mathrm{ISM}$. 

The quantity that correlates most with  $\emy$ and $\emyrat$ is $\Sigma_\star$: the non-parametric Spearman's test gives a correlation coefficient $\rho=0.93$ for $\Sigma_\star$ versus $\emy$ and $\rho=0.89$ for $\Sigma_\star$ versus $\emyrat$ (in all cases discussed here, the correlations are significant at a confidence level $>$ 99.99\%). As a comparison, $\emy$ and $\emyrat$ are correlated among themselves with $\rho=0.86$. Besides being used in Eq.~\ref{eq:b13}, $\Sigma_\star$ can also be considered as a proxy for the optical-NIR ISRF: thus, dust in regions with higher $\Sigma_\star$ have higher $\emyrat$ and $\emy$. Instead, the UV ISRF appears to contribute less to the dust heating. This is shown by the anticorrelation between $S_\mathrm{NUV}/S_\mathrm{3.6\,\mu m}$ and $\emy$, with  $\rho=-0.55$ ($\rho=-0.67$
 for $\emyrat$). We found that one galaxy, NGC~3521, characterized by a low global $S_\mathrm{NUV}/S_\mathrm{3.6\,\mu m}$ ($\approx$ 0.02, using data from \citealt{ClarkA&A2018}) also has low $\emy$ for $\emyrat$=5-6 (see the left panel of Fig.~\ref{fig:emy250_xcob}). Pixels from this galaxy are responsible for lowering slightly the average $S_\mathrm{NUV}/S_\mathrm{3.6\,\mu m}$ in Fig.~\ref{fig:emy250_xcob_cmaps}, on the  low-emissivity side of the outer contour plot. However, this does not seem to be a general property of other galaxies of lower UV emission: for example, NGC~4736 and NGC~5055, with global  $S_\mathrm{NUV}/S_\mathrm{3.6\,\mu m} \approx 0.03$,  and NGC~5194, with $S_\mathrm{NUV}/S_\mathrm{3.6\,\mu m} \approx 0.06$, follow the mean trend with the other galaxies (all with global $S_\mathrm{NUV}/S_\mathrm{3.6\,\mu m} \ga 0.1$).

Strong correlations are also found for $f_\mathrm{mol}$, with Spearman's $\rho=0.81$ with respect to $\emy$ and $\rho=0.85$ with respect to $\emyrat$. 
For $\emy$ versus $f_\mathrm{mol}$, the correlation could, in principle, be boosted by the fact that both quantities depend on $1/N_\mathrm{H}$; yet the errors on gas
column density are small and indeed a test shows a null correlation if we assume that the $250\, \mu$m surface density is completely uncorrelated to
the gas column density (we used the same random procedure as for the "null" test, this time by shuffling the pixel values for the gas column density).
The enhancement of emissivity in regions with higher molecular gas fraction is in agreement with \citet{BianchiA&A2019}, where the same result is found for mean galactic properties. This behavior translates, with some scatter and a reduction in the correlation, to that for $\Sigma_\mathrm{ISM}$, where it is $\rho=0.52$ with respect to $\emy$ and $\rho=0.63$ with respect to $\emyrat$. 
In general,  $\Sigma_\mathrm{ISM}$  is larger for higher $\emyrat$ values. 
For $\emyrat < 5$, there also seems to be a trend of higher $\emy$  for pixels with a lower column density.  However, this is a spurious effect due to the dependence of $\emy$ on $1/N_\mathrm{H}$: in this range, the "noisier" pixels in $\emyrat$ determine a roughly constant lower limit in $I(250 \mu\mathrm{m})$, regardless of their gas column density; thus $\emy$ is boosted in pixels of lower $\Sigma_\mathrm{ISM}$ (belonging mostly to NGC~5457). 

Overall, $\emyrat$ and $\emy$ are higher in the central regions of a galaxy, where the stellar and total gas surface density, as well as the molecular gas contribution to the ISM
(along with metallicity, for which we used a smooth negative gradient) are high, and the UV radiation field is reduced with respect to the optical-NIR.  
Indeed, for each galaxy, strong inverse correlations are found with the galactocentric distance, $d$: for example, the Spearman's correlation coefficient for  $\emy$ versus $d$ ranges from 
$\rho=-0.83$ for NGC~5194 to $-0.97$ for NGC~5055. When the whole sample is considered, and the distances scaled on the optical radii of each galaxy, the correlation is reduced, with $\rho=-0.61$
for $\emy$ versus $d/R_{25}$. This may be due to the fact that a simple scaling on $R_{25}$ is not able to describe the relevant gradients in physical conditions. In fact, when scaling the distance on the stellar scalelength 
$h_\star$ from \citet{CasasolaA&A2017}, which at least captures the variation in the stellar surface density of the disk (but not of the central parts of a galaxy), the correlation for 
$\emy$ versus $d/h_\star$ is raised to $\rho=-0.73$. In any case, the enhancement of $\emy$  with radiation field and gas density is dominated by the smooth galactic gradient.
A possible effect  along spiral arms is instead not evident in our analysis,  because of their smaller dynamic range
and because of the low resolution imposed by the 500 $\mu$m data, resulting in the lack of an evident trace of spiral arms in the
$\emyrat$ maps (as shown by Fig.~\ref{fig:images} and \ref{fig:otherimages}).

\section{Discussion}
\label{sec:discussion}

\begin{figure}
\includegraphics[scale=0.3]{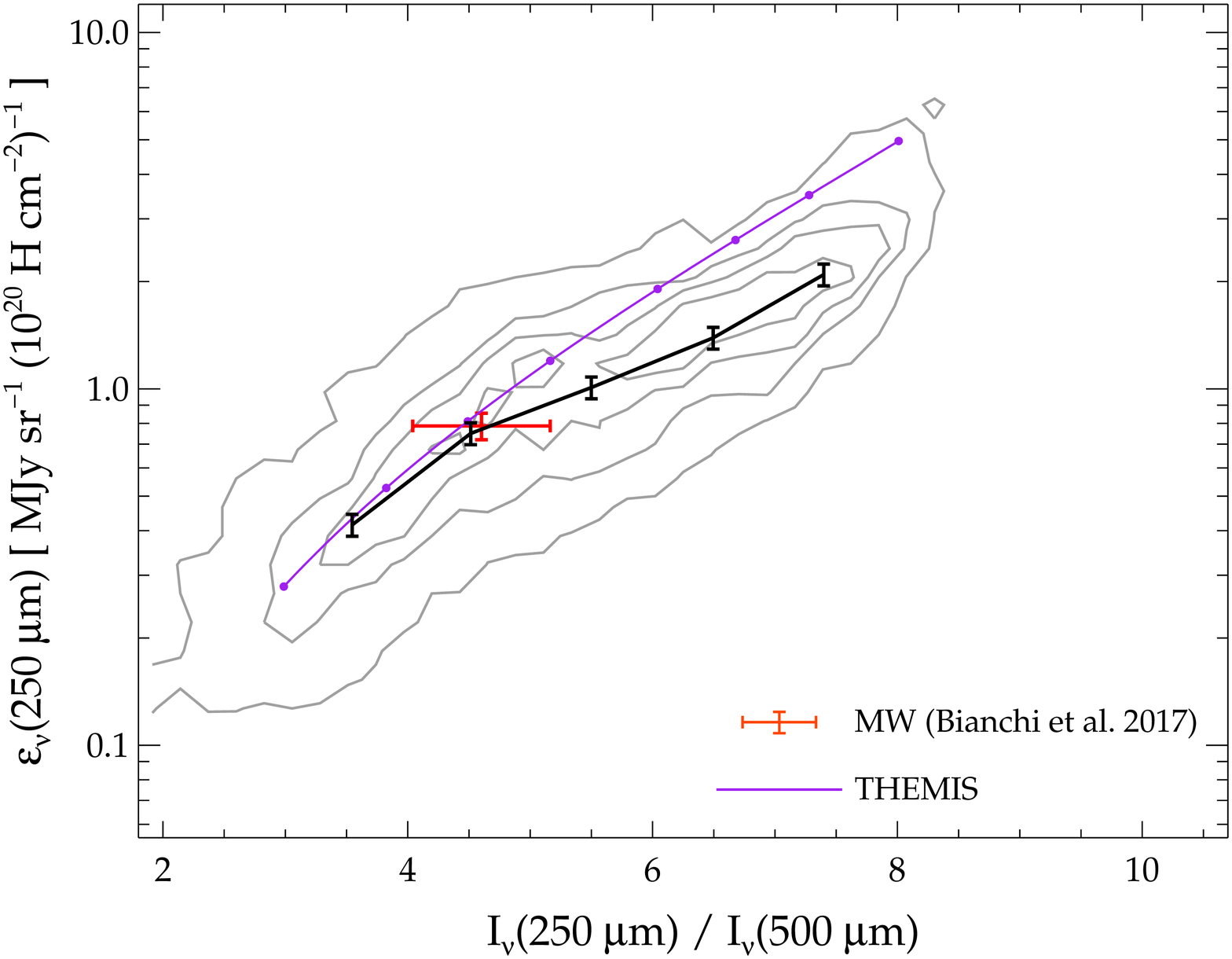}
\caption{
$\emy$ versus  $\emyrat$. Comparison between our results and the MW value for the cirrus. We also show the behavior of the THEMIS dust model under ISRFs of growing intensities  (dots are for U= 0.2, 0.5, 1, 2, 5, 10, 20, and 50 from left to right, with ISRF=U$\times$LISRF; see Appendix~\ref{app:bias}). Grey contours and average emissivity as in Fig.~\ref{fig:emy250_xcob}. The errorbars for the average emissivity show the impact on $\emy$ of the SPIRE calibration error.
}
\label{fig:emy250_xco_b13_mw}
\end{figure}

The average emissivity in the $\emyrat = 4.5$ bin is very close to the value found for the high-latitude Galactic cirrus,  $\emy = 0.79\pm0.07$ MJy sr$^{-1}$ (10$^{20}$ H cm$^{-2}$)$^{-1}$ for $\emyrat=4.6\pm0.6$ (Fig.~\ref{fig:emy250_xco_b13_mw}; see \citealt{BianchiA&A2017,BianchiA&A2019} for the MW 
estimate\footnote{The uncertainties given here are larger than those of \citet{BianchiA&A2019}, because we now include the
calibration error (and correct an error in the previous estimate of the $\emyrat$ uncertainty). While the calibration error is not relevant 
in the comparison of data obtained with the same instrument at the same wavelength (like the SPIRE determinations of $\emy$ done in this work and that for the MW in
\citealt{BianchiA&A2017}), it should be considered when comparing to independent estimates (e.g.,\  the $\emy$ predictions from THEMIS in Fig.~\ref{fig:emy250_xco_b13_mw})
and to estimates at different wavelengths obtained with different instruments and calibration procedures (e.g.,\ $\epsilon_\nu$  discussed later in Sect.~\ref{sec:spectrum}).}). This reassures on the use of the MW emissivity to derive dust masses in other galaxies:
as shown in the previous section, the intrinsic scatter around this value is limited to $35\%$. The scatter possibly hides variations in the dust properties within a galaxy and from object to object, even though we could find no  clear dependence on any measurable quantity.

For the expected values of $\emy$ at different $\emyrat,$ we have to rely on dust grain models. In Fig.~\ref{fig:emy250_xco_b13_mw}, we show the behavior of the THEMIS model for diffuse dust under varying intensities of the ISRF heating the grains (from less to more intense ISRFs, from left to right; see Appendix~\ref{app:bias} for details on the calculation). For  
$\emyrat \la 5$ the dust model follows the average trend of the data, within the SPIRE $250 \,\mu$m calibration uncertainty.  
Indeed, THEMIS is made to pass from the MW emissivity estimate, where the heating field is the LISRF (the dot with $U=1$ in Fig.~\ref{fig:emy250_xco_b13_mw}). However, for $\emyrat \ga 5$ the THEMIS prediction overestimates the average emissivity: for $\emyrat=$ 5.5, 6.5, 7.5, the average $\emy$ is a factor 1.4, 1.7 and 1.9 lower, respectively, than the expectation for the model at the same color ratios, while we found a total scatter of 30\% only (including photometric uncertainties). A similar discrepancy is found in \citet{BianchiA&A2019}.

The most naive interpretation of the tension between our estimates and the theoretical predictions is that the lower $\emy$ for higher $\emyrat$ is simply due to a reduction in the dust absorption cross section with respect to that of THEMIS. If this is true, the dust mass surface density for pixels with $\emyrat \ga 5$ could be underestimated by up to a
factor $\approx 2$, when using the dust properties from the model. The impact of this underestimation on the global dust mass in a galaxy would depend on the fraction of the dust emission at higher $\emyrat$. We tested this by using the dust maps obtained using THEMIS in \citet{CasasolaA&A2017} and raising the dust surface density for the pixels in the three higher $\emyrat$ bins, by the average factors discussed in the previous paragraph. For two galaxies, the total dust mass after this correction is within 10\% of the original estimate: they are NGC~925 and NGC~2403 which (as we have seen) have a dearth of pixels with $\emyrat > 5$. For the other galaxies, the corrected dust mass would be from 35 to 55\% higher, with
the exception of NGC~5194, where it would rise by 75\%.

However, there are other possible explanations for the lower-that-expected $\emy$ at higher $\emyrat$, apart from global variations in the dust properties. We explore them in the rest of this section.

\subsection{ISRF mixing}

The predictions of Fig.~\ref{fig:emy250_xco_b13_mw} rely on the simplifying assumption that all of the dust is heated by a homogeneous ISRF. However, along any line of sight (i.e., the pixel level in our analysis) dust might be subject to a range of heating conditions with, for example, most of the grains absorbing light from a less intense ISRF, and a fraction of them being  heated to higher temperatures in a more intense field. Indeed, when a dust grain mixture is exposed to a distribution of ISRFs, the resulting spectrum is broader: this might alter the position and trend of the model  predictions in the $\emy$-$\emyrat$ plane.

We explore the effects of  this mixing of heating conditions in Appendix~\ref{app:bias}, by testing the ISRF distributions which are typically used to fit the global or resolved SED of galaxies. In general, we found that the effects of ISRF mixing are small, at least when emission at longer wavelength is concerned: the trends we found are not much different from the case of homogeneous ISRF shown in Fig.~\ref{fig:emy250_xco_b13_mw}. A moderate bias could be present for $\emyrat = 4.5$, with $\emy$  $\sim$10\% lower (at maximum) than the values from a single intensity ISRF. However, the difference reduces for higher $\emyrat$ and cannot explain the lower values for  $\emy$ we find.

Another simplification in the modeling of dust heating is to assume always the same spectral shape for starlight, namely, that of the LISRF. Instead, dust could be heated by UV-dominated ISRFs near star-forming regions or by a redder spectrum in the diffuse medium close to the center of galaxies. Nevertheless, we found that the use of different spectra does not  significantly change the modeled trends in the $\emy$-$\emyrat$ plane (Appendix~\ref{app:bias}).

\subsection{Dust in different environments}

\citet{ClarkMNRAS2019} used {\em Herschel} maps of two galaxies (one of which, NGC~628, is also included in our sample) to derive the dust absorption cross section, after 
having estimated the total dust mass from gas and metallicity measurements. They reported a reduction of  the dust absorption cross-section
for lines of sight of higher $\Sigma_\mathrm{ISM}$. While we do not derive the absorption cross-section in this work  and limit our analysis to the
emissivity only, we support a similar view: pixels with $\emyrat > 5$, where the average  $\Sigma_\mathrm{ISM}$ is higher (see Fig.~\ref{fig:sism}), have a 
reduced $\emy,$ with respect to model predictions.  

In dense environments  dust grains are expected to accrete atoms from the ISM, develop ice mantles, and coagulate: all these processes tend to increase the absorption cross-section (and emissivity) rather than decrease it (see, e.g., \citealt{KoehlerA&A2015}). In \citet{BianchiA&A2019} we tested the results of  \citet{KoehlerA&A2015} and confirmed that, even for the less evolved grain mixture studied in that work (the CMM model), $\emy$ is higher than what expected for diffuse dust. Furthermore, the evolution of dust grain properties is supposed to occur in regions where the ISRF is shielded, and thus at lower $\emyrat$, while we find a discrepancy with the diffuse dust model at larger $\emyrat$.

\citet{PriestleyMNRAS2020} suggest that the results of  \citet{ClarkMNRAS2019}  are biased: along the lines of sight with higher column density, emission from dust exposed to the 
ISRF in the diffuse medium could be mixed with emission coming from dense molecular clouds; here, grains  attain a lower temperature because they are reached by a lower 
intensity ISRF, attenuated by extinction from dust itself. Neglecting the presence of colder dust and using a single temperature to model the SED, as done by 
\citet{ClarkMNRAS2019},  results in a lower estimate of the absorption cross sections, essentially because colder dust contributes only minimally to the overall emission
(while it is accounted for in the dust mass estimated from the gas metallicity). In Appendix~\ref{app:bias}, we tested the impact on our estimates of the scenario  proposed by
 \citet{PriestleyMNRAS2020}. We found that $\emy$ could be biased low by a significant amount only for the higher $\Sigma_\mathrm{ISM}$ we found
in our sample, and in particular for lower intensity ISRFs. Instead, the bias is smaller for the more intense ISRFs that should be responsible for the emission at larger $\emyrat$.
In fact, even assuming different properties for dust in denser environments (the CMM model of \citealt{KoehlerA&A2015}), the trend of $\emy$ versus $\emyrat$ is not substantially altered from that predicted for single-ISRF models. The bias described by \citet{PriestleyMNRAS2020} cannot thus explain the  discrepancy between 
model expectations and $\emy$ estimates at $\emyrat > 5$.

\subsection{Dependence on $X_\mathrm{CO}$}

\begin{figure}
\includegraphics[scale=0.3]{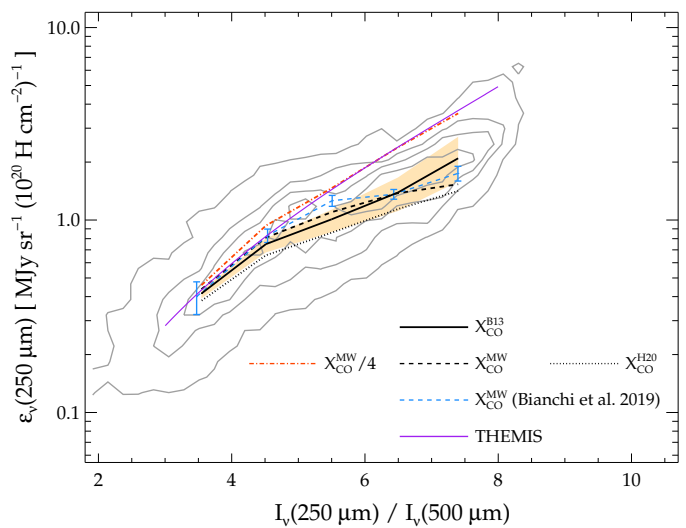}
\caption{
Dependence of the average  $\emy$ versus  $\emyrat$ on  $X_\mathrm{CO}$. The shaded area shows the uncertainty due to the CO-to-H$_2$ conversion. Contours and average emissivity as in Fig.~\ref{fig:emy250_xcob}, model predictions as in Fig.~\ref{fig:emy250_xco_b13_mw}. We also show the results of \citet{BianchiA&A2019} for $X_\mathrm{CO}^\mathrm{MW}$.
}
\label{fig:emy250_xco_b13}
\end{figure}

So far, we only considered the uncertainties due to photometry. However, there are also large uncertainties in the CO-to-H$_2$ conversion factor. We estimated them for  
$X_\mathrm{CO}^\mathrm{B13}$, using a Monte Carlo procedure and building random representations of the $\emy$ versus $\emyrat$ distribution according to the scatter in
$X_\mathrm{CO}^\mathrm{MW}$ and $R_{21}$ described in Sect.~\ref{sec:method} (and to the scatter in the metallicity gradients from \citealt{PilyuginAJ2014}, although
this is a minor effect; see also Appendix~\ref{app:metgrads} for the impact of different metallicity calibrators). Since the uncertainty in $X_\mathrm{CO}^\mathrm{MW}$ is 
common to all galaxies, we draw a single random value for this quantity in each representation, while we draw independent values of $R_{21}$ for each galaxy (thus assuming 
that the measured scatter is due to galaxy-to-galaxy variations; we used the estimate of \citealt{LeroyApJ2022}). After running a thousand representations, we measured the 
standard deviation of the random $\emy$ trends, shown by the shaded area in Fig.~\ref{fig:emy250_xco_b13}. 
At  $\emyrat = 4.5,$ the uncertainty due to $X_\mathrm{CO}^\mathrm{B13}$ is 8\%, much smaller than the scatter in the dataset: this is the reflection of the very low 
contribution of molecular gas to $N_\mathrm{H}$ and thus on the little dependence of $\emy$ on $X_\mathrm{CO}$ for the first $\emyrat$ bins. 
For larger $\emyrat$, instead, the contribution of molecular gas increases (see $f_\mathrm{mol}$ in Fig.~\ref{fig:emy250_xcob_cmaps}) and
the uncertainty in $\emy$ due to the CO-to-H$_2$ conversion factor grows: at $\emyrat=7.5$ it is 30\%, the same as the scatter in the observations. 
Even combining this uncertainty with the data scatter, the difference between the observed $\emy$ and the model cannot be explained.

\subsubsection{Different $X_\mathrm{CO}$ recipes}

The $\emy$ versus $\emyrat$ trend depends, though not strongly, on the recipe adopted for $X_\mathrm{CO}$. For our reference choice, we have an average 
$X_\mathrm{CO}^\mathrm{B13}$ = 1.5$\times  X_\mathrm{CO}^\mathrm{MW}$ for the  $\emyrat=4.5$ bin (the value being increased by the low metallicity of the
contributing pixels, Z $\approx 0.5$ Z$_\odot$). For $\emyrat = 7.5$, the average surface density in our sample is 
$\Sigma_\mathrm{total} \approx$ 300 M$_\odot$ pc$^{-2}$ and the metallicity close to Solar, Z $\approx 0.9$ Z$_\odot$
(those are also the pixels closer to the galactic centers). For these values, the average conversion factor reduced to 
$X_\mathrm{CO}^\mathrm{B13} \approx 0.6 \times X_\mathrm{CO}^\mathrm{MW}$, making the CO-brighter regions contribute less to 
the molecular gas column density. This is evident, for example, when comparing the two $N_\mathrm{H_2}$ maps for NGC~5457 in Fig.~\ref{fig:images}: 
the map obtained assuming the constant $X_\mathrm{CO}^\mathrm{MW}$ throughout the galaxy -- resulting from a simple rescaling  
of the original CO map -- has a central peak which lacks in the map for $X_\mathrm{CO}^\mathrm{B13}$ (where the conversion factor is smaller in the center).
As a result, $\emy$ increases in the central parts of NGC~5457 (and, in general, for the regions in all galaxies where  $\emyrat$ is higher). This increase, however,
is not sufficient to reach the model predictions, as already discussed. The emissivity can be boosted further by lowering the $\Sigma_\mathrm{total}$ threshold in Eq.~\ref{eq:b13}. However, even reducing it to half of the reference value, the trend remains still within the uncertainties estimated for $X_\mathrm{CO}^\mathrm{B13}$. We also tested a  CO-to-H$_2$ conversion factor including possible variations of $R_{21}$ with the environment: we used the dependence on  $I_\mathrm{CO(2-1)}$ discussed by \citet{BrokMNRAS2021}, but the trend differs from the reference value only by a few percent (not shown).

Despite the dependence on metallicity and surface density, the values obtained for the $X_\mathrm{CO}^\mathrm{B13}$ recipe do not differ much from, 
and encompass, those typically used in the MW. As a consequence, the trend for $\emy$ obtained assuming $X_\mathrm{CO}^\mathrm{B13}$ is similar to that 
for a constant $X_\mathrm{CO}^\mathrm{MW}$  (the average is shown as a dashed black line in Fig.~\ref{fig:emy250_xco_b13}).  
In Fig.~\ref{fig:emy250_xco_b13} we also
show the average trend obtained by \citet{BianchiA&A2019} under the same assumption: that result for the average properties 
in a sample of 204 galaxies is very similar to (and in some bins entirely consistent with) what we find here on resolved pixels for nine objects.

\begin {table*}
\caption{Average emissivities $\epsilon_\nu$ and standard deviations, in units of MJy sr$^{-1}$ (10$^{20}$ H cm$^{-2}$)$^{-1}$, for all wavelengths and $\emyrat$ bins.
The $X_\mathrm{CO}^\mathrm{B13}$ conversion factor is assumed.}              
\label{tab:emy}      
\small
\centering                                      
\begin{tabular}{c c c c c c c c c }          
\hline\hline                        
$\langle\frac{I_\nu(250 \,\mu\mathrm{m})}{I_\nu(500 \,\mu\mathrm{m})}\rangle$ & 
$70 \,\mu\mathrm{m}$ & $100 \,\mu\mathrm{m}$& $160 \,\mu\mathrm{m}$ & $250 \,\mu\mathrm{m}$ & $350 \,\mu\mathrm{m}$ & $500 \,\mu\mathrm{m}$ & $N_\mathrm{pixels}^{(a)}$ \\
\hline                                   
3.5 &  0.21$\pm$0.17 &   0.38$\pm$0.25 &   0.58$\pm$0.31 &   0.42$\pm$0.19 &   0.24$\pm$0.10 &   0.12$\pm$0.05 &  2228 (  2034 ) \\
4.5 &  0.45$\pm$0.31 &   0.78$\pm$0.49 &   1.19$\pm$0.58 &   0.75$\pm$0.30 &   0.39$\pm$0.15 &   0.17$\pm$0.06 &  2933 (  2323 ) \\
5.5 &  0.59$\pm$0.42 &   1.10$\pm$0.62 &   1.75$\pm$0.80 &   1.01$\pm$0.35 &   0.48$\pm$0.16 &   0.18$\pm$0.06 &  2771 (  2301 ) \\
6.5 &  0.93$\pm$0.46 &   2.00$\pm$0.85 &   2.66$\pm$0.96 &   1.39$\pm$0.41 &   0.60$\pm$0.16 &   0.21$\pm$0.06 &  2733 (  2337 ) \\
7.5 &  1.87$\pm$0.86 &   3.70$\pm$1.46 &   4.47$\pm$1.55 &   2.09$\pm$0.65 &   0.84$\pm$0.25 &   0.28$\pm$0.08 &  1905 (  1373 ) \\
\hline                                             
\end{tabular}
\tablefoot{
\tablefoottext{a}{Number of pixels in each $\emyrat$ bin (in parenthesis the reduced number of pixels available at 100 $\mu$m).}
}
\end{table*}

In Fig.~\ref{fig:emy250_xco_b13}, we also show the trend obtained using $X_\mathrm{CO}^\mathrm{H20}$  (dotted line). This recipe, based
on global galactic values,  has a stronger dependence on metallicity, possibly driven by the relatively smaller number of low-metallicity galaxies in the 
sample of \citet{HuntA&A2020}; thus, it is probably not suitable for our resolved analysis of relatively higher-metallicity objects. Nevertheless,
the trend  is not much different from the rest; the exception is that it has lower $\emy$, since the average conversion factors are larger: $X_\mathrm{CO}^\mathrm{H20}
\approx 3 \times X_\mathrm{CO}^\mathrm{MW}$ and $1.2 \times X_\mathrm{CO}^\mathrm{MW}$, at $\emyrat=4.5,$ and 7.5, respectively.
The $X_\mathrm{CO}^\mathrm{H20}$ recipe has the advantage of resulting in a smaller scatter of $\emy$ in the highest $\emyrat$ bin: it is 22\%.
However, the scatter produced by the simpler constant $X_\mathrm{CO}^\mathrm{MW}$ recipe is even smaller, 15\%.
Thus, neither $X_\mathrm{CO}^\mathrm{H20}$ nor $X_\mathrm{CO}^\mathrm{B13}$ reduce the scatter with respect to the constant value, 
as would naively be expected if the adopted formulas were to catch all possible variations of the conversion factor (and dust properties do not change 
significantly within a galaxy and across galaxies).

\subsubsection{Dust-based $X_\mathrm{CO}$ estimates}

If the dust properties, relative to the total gas content, do not vary between galaxies and across environments, it is tempting to use the dust mass or surface density as a tracer of the total  ISM  (atomic and molecular) budget \citep{EalesApJ2012, SandstromApJ2013, JiaoMNRAS2021}. This assumption is used, for example, by \citet{SandstromApJ2013}, to derive the 
$X_\mathrm{CO}$ values and gas-to-dust ratios (and their variations) along the disk of 26 nearby galaxies. Basically, these authors assumed that these quantities do not vary substantially over scales smaller than 1 kpc. On this scale, their solution requires that the total surface density of the gas, obtained by multiplying the dust mass surface density by the gas-to-dust mass ratio, must equal the sum of HI  and molecular gas surface densities -- with the latter derived from CO observations converted via the  $X_\mathrm{CO}$ factor (and $R_\mathrm{21}$). The method obviously relies on the independent determination of the dust mass via a MW dust model (that of  \citealt{DraineApJ2007b}). \citet{SandstromApJ2013} found $X_\mathrm{CO}$ values
that are slightly lower, but still compatible, with the MW value, with little variation across most of the disks. It is only in galactic centers that $X_\mathrm{CO}$ is reduced with respect to the disk average. A similar analysis done by  \citet{JiaoMNRAS2021} has offered analogous conclusions.

Our findings apparently suggest that the model dust emissivity is overestimated in denser environments, where the contribution of molecular gas to the total surface density is higher.
If this is true, the model-dependent dust (and total gas) content  would be underestimated: the lower $X_\mathrm{CO}$ found by \citet{SandstromApJ2013} and \citet{JiaoMNRAS2021}  might potentially be due to this bias. It is true that both papers partially consider variations in the dust properties, by allowing 
the dust-to-gas mass ratio to vary with respect to the value that is intrinsic to the model grain mixture. However, this approach only scales the amount of dust grains along different sightlines, but does not allow for variations in the emission properties as a function of wavelength. Another problem might arise from the use of the \citet{DraineApJ2007b} model,
whose emissivity does not match that of the MW cirrus: nevertheless, this global variation can be taken into account in the fitted gas-to-dust mass ratio, while the $\emy$ versus $\emyrat$ trend of the dust model is similar to that of THEMIS  \citep{BianchiA&A2019}.
 
Since  the $X_\mathrm{CO}^\mathrm{B13}$  recipe adopted in this work is in part derived from the results of \citet{SandstromApJ2013},
we might also
incur the paradox of studying dust emission properties derived from a conversion factor that is itself based on the assumption of a dust model (albeit a different one). Nevertheless, we do not believe that our findings are 
affected by this choice. In fact, the $X_\mathrm{CO}^\mathrm{B13}$ recipe is also based on the dust-independent measurements for $X_\mathrm{CO}^\mathrm{MW}$, 
and on gas spectral line modeling for denser regions, such as those in ultraluminous IR Galaxies. The reduction of the conversion factor in the inner parts of galaxies was also recently confirmed by \citet{IsraelA&A2020}, where values as low as $0.1\times X_\mathrm{CO}^\mathrm{MW}$ in galactic central cores were found from the line modeling. 
Finally, it is to be noted that, apart from the last $\emyrat$ bin, the average trend of Fig.~\ref{fig:emy250_xco_b13} 
is maintained also using the constant $X_\mathrm{CO}^\mathrm{MW}$ value. 

Matching the emissivity trend to the model predictions indeed requires a smaller $X_\mathrm{CO}$, not only in galactic nuclei but over the whole disk.
The average emissivity can be aligned to the dust model by using a constant  $X_\mathrm{CO} = 0.25 \times X_\mathrm{CO}^\mathrm{MW}$ 
for all pixels,  regardless of metallicity or surface density (red dot-dashed in Fig.~\ref{fig:emy250_xco_b13}). 
Since this conversion factor is
significantly different from the values normally accepted for this quantity, it seems reasonable to conclude that the lower than expected $\emy$ at higher $\emyrat$
is not due to an erroneous choice for  $X_\mathrm{CO}$  (or produced by the other biases discussed in this section),  -- but this is, rather, the  result of actual variations in the 
dust properties and, in particular, of a reduction in the absorption cross-section.

\section{Emissivity SED}
\label{sec:spectrum}

\begin{figure*}
\sidecaption
\includegraphics[width=12cm]{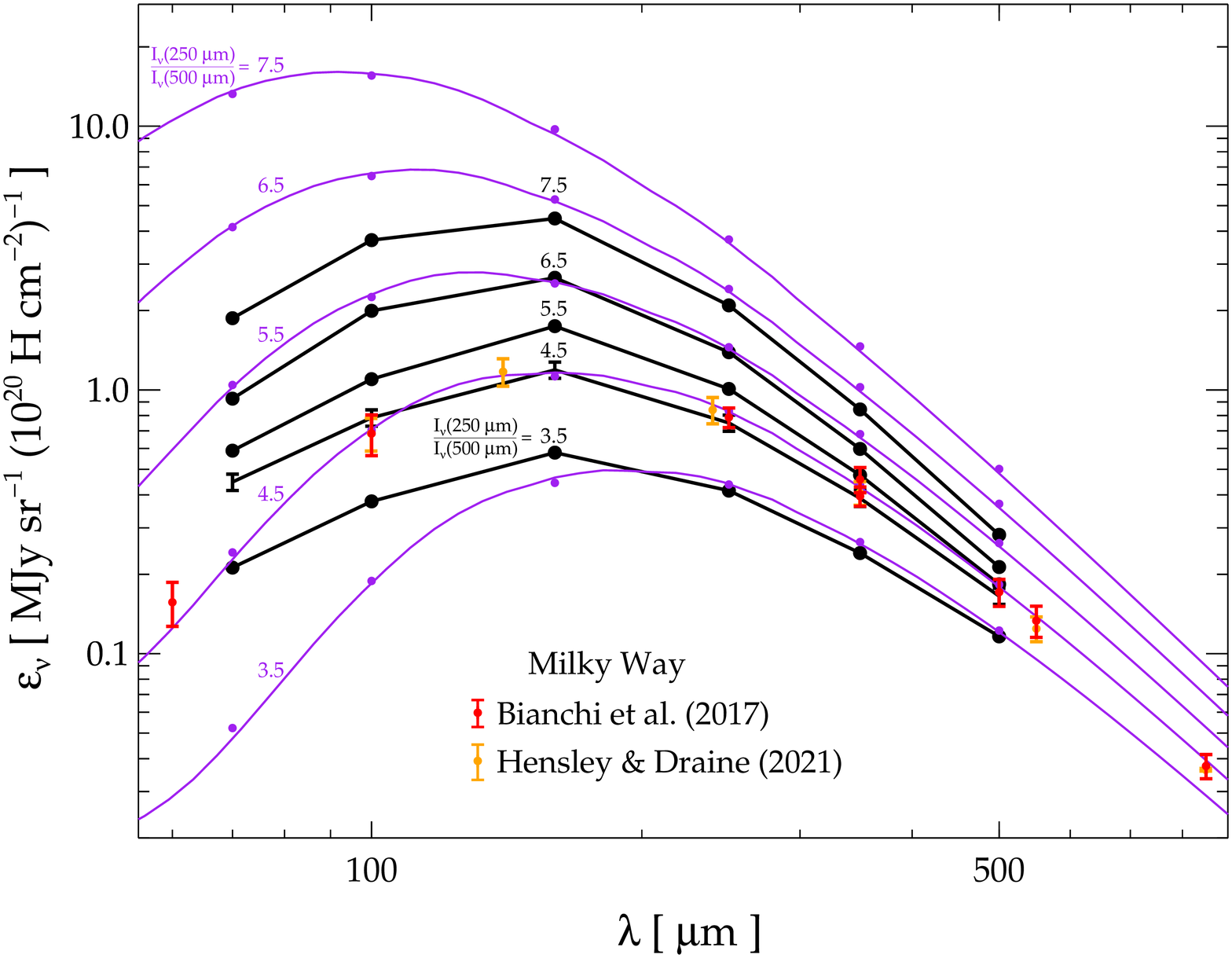}
\caption{
Average emissivity SED for the five $\emyrat$ bins of Fig.~\ref{fig:emy250_xcob}, using $X_\mathrm{CO}^\mathrm{B13}$ (black lines). For $\emyrat = 4.5$, the calibration errors are shown.  Purple lines show the SED from the THEMIS model, averaged over the $\emyrat$ values in each bin. Purple dots mark the emissivities after averaging over the {\em Herschel} filter response functions. The MW
emissivities are also shown.
}
\label{fig:emy_sed_xcovgpb}
\end{figure*}

Using Eq.~\ref{eq:emy}, we also derived the emissivity in the other {\em Herschel} bands. The average emissivities and the standard deviation for all wavelengths and 
$\emyrat$ bins are given in Table~\ref{tab:emy}. The same number of pixels is available for each band at each $\emyrat$ bin. The only exception is 100 $\mu$m, where the data for  
NGC~2403 and NGC~5194 are not available. Nevertheless, the available pixels in this band are 82\% of those available for the rest of the spectral coverage, thus a bias is unlikely.  
The average SED for each  $\emyrat$ bin is shown in Fig.~\ref{fig:emy_sed_xcovgpb}. As for the average $\emy$ in Fig.~\ref{fig:emy250_xcob}, the standard deviation of the mean is smaller than the symbols. For the $\emyrat=4.5$ case we show, as a reference, the total calibration uncertainties for the PACS and SPIRE bands (about 7\% and 6\%, respectively, using the values quoted by \citealt{GallianoA&A2021} and assuming for PACS the same uncertainties for extended sources as for SPIRE).

At $\emyrat = 4.5$, the emissivities we estimated for  $\lambda\ga 100\, \mu$m are all compatible with the observations in the MW high-latitude cirrus, and with the predictions of  the 
THEMIS diffuse dust model, which are set to fit those observations (see the datapoints and the second purple line from the bottom in Fig.~\ref{fig:emy_sed_xcovgpb}).
At 70 $\mu$m, instead, the emissivity is higher that what predicted by THEMIS: this is expected on the basis of the stronger bias due to ISRF mixing at shorter wavelengths, while the cirrus emission likely results from dust heated by a single radiation field, the LISRF. As discussed in  Appendix~\ref{app:bias}, for the wavelength range considered here, ISRF mixing results in a stronger bias when the radiation fields have lower intensities: this is shown by the SED for  $\emyrat  =  3.5$, where our estimates are larger than model predictions
up to $\lambda\la 160 \mu$m; however, they are still close to the corresponding THEMIS model for $\lambda \ge 250 \mu$m.

For $\emyrat > 5$, we see the same trends as for $\emy$ in all the {\em Herschel} bands, with the emissivity SED being systematically lower than model predictions
(compare the black and purple lines for the same $\emyrat$ value in Fig.~\ref{fig:emy_sed_xcovgpb}).  For $\lambda\ge 160\mu$m the THEMIS model could match the average SEDs if scaled by a factor 1/1.4 for the $\emyrat  = 5.5$ bin and 1/1.9 for the $\emyrat > 6$ bins (i.e., the ratios at $250 \, \mu$m  discussed in Sect.~\ref{sec:discussion}). 
In principle, we could made the THEMIS model reproduce the lower $\emy$ by lowering its dust-to-gas ratio (0.0074 for the original dust mixture; \citealt{JonesA&A2017}) by the same
factors, while retaining the size distribution and composition (and thus the spectral shape of dust emission). However, this cannot be a plausible explanation: the pixels in the larger  $\emyrat$ bins have a larger metallicity than in the $\emyrat = 4.5$ bin (the "MW reference"), being closer to the galactic centers; thus, their dust-to-gas ratio should be higher, rather than smaller, than for the THEMIS model. Furthermore, for   $\lambda< 160\mu$m, the average emissivity SED has a different spectral shape than models: the ratio $I_\nu(70 \,\mu\mathrm{m})/I_\nu(160 \,\mu\mathrm{m})$, for example, should be higher than in the corresponding model, because of ISRF mixing (or tend to the model, for the more intense ISRFs; see Appendix~\ref{app:bias}); instead, it is lower. The incompatibility of the rescaling factor with the growth of dust-to-gas ratio with metallicity, and the overall difference in spectral shape of the average emissivities with respect to the model, suggests that there could be a variation in the properties of the dust mixture (either in the relative ratios of the various components or in the size distributions, or both).

\section{Summary and conclusions}
\label{sec:summary}

We derived the FIR/submm dust emissivity on a resolved scale ($\approx$ 400-500 pc) for a sample of nine low-inclination, nearby spiral galaxies. We used maps of 
the dust emission from the {\em Herschel} satellite as well as maps of the atomic and molecular gas column density from the THINGS and HERACLES surveys. Our main findings are as follows:
\begin{enumerate}
\item The average $\emy$ of our sample is compatible with estimates in the MW, for  $\emyrat = 4.5$ (i.e., the color of dust emission in the MW cirrus). The scatter 
around this value is limited to 35\%, with no evident dependence on local or global galactic properties.
\item For $\emyrat > 5$, the average emissivity is up to a factor $\sim$2 lower than what is expected based on predictions from models of MW dust. 
The pixels showing these colors are, on average, closer to the galactic centers and have a higher stellar and gas column density, as well as a higher molecular gas fraction.
\item The results do not depend strongly on the main assumption of the analysis,
namely, the choice for the $X_\mathrm{CO}$ factor;
rather, they can be reconciled with the predictions of dust models only using 
$X_\mathrm{CO}$ values that are much smaller than what is generally found in the MW and other spirals.
\item 
The same conclusions are valid for the emissivity $\epsilon_\nu$ in all {\em Herschel} bands with $\lambda\ge 100\, \mu$m.
\end{enumerate}

Our resolved analysis confirms the results obtained by \citet{BianchiA&A2019}, where the disk-averaged $\emy$ was derived based on a sample of 204 objects. 
Unless the actual $X_\mathrm{CO}$ factor used to derive the molecular gas surface density is significantly different from the recipes currently available in the literature,
the trends we find for $\epsilon_\nu$ versus $\emyrat$ suggest variations of the dust properties in regions where the radiation field is stronger than the LISFR.

Typically, dust masses in extragalactic objects are derived using grain models based on the MW high latitude cirrus emissivity, which is matched by our extragalactic 
estimates for $\emyrat = 4.5$. As we discuss in Sect.~\ref{sec:discussion}, adopting these models to derive the dust masses (or surface densities) for $\emyrat > 5$ 
could imply an underestimation by up to a factor of two. This uncertainty in the dust mass determination adds to other uncertainties that have been pointed out, such as the comparable
discrepancies between dust mass estimates obtained using different dust models \citep{NersesianA&A2019,GallianoA&A2021} and the generally poor knowledge of the
emission properties of cosmic dust analogues: for instance, laboratory measurements of the absorption cross sections for amorphous silicates at FIR/submm wavelengths 
have recently questioned the validity of the optical properties used in current dust models \citep{DemykA&A2017} 
and have been claimed to result in dust mass overestimates by factors ranging from 2 to 20  \citep{FanciulloMNRAS2020}.

As a final note, we recall that our analysis assumes homogeneous dust and gas conditions within a few hundred parsecs. While we explored the bias due  to a range of starlight intensities, we cannot exclude that the estimated emissivity, even within our spatial scale, could result from dust emission coming from different regions, with different grain properties, gas column 
densities, and $X_\mathrm{CO}$ conversion factors. These further biases might affect not only the determination of the emissivity, but also of the dust mass surface density (once the emissivity is assumed from grain models) and $X_\mathrm{CO}$ determinations (either when the dust mass is used as a proxy for the gas or otherwise); they could be tested numerically with high resolution radiative transfer simulations (as in, e.g., those of \citealt{KapoorMNRAS2021} and \citealt{CampsMNRAS2022}), taking into account the evolution of dust properties with the environment; they could also be checked observationally when higher resolution data for the diffuse continuum emission become available.
 
\begin{acknowledgements}
We dedicate this paper to Jonathan Ivor Davies (1955-2021), friend, colleague, and leader of the DustPedia collaboration.
We acknowledge support from the INAF mainstream 2018 program "Gas-DustPedia: A definitive view of the ISM in the Local Universe" and from grant PRIN MIUR 2017- 20173ML3WW\_001. We thank E. M. Xilouris for useful comments.
\end{acknowledgements}

\bibliographystyle{aa} 
\bibliography{/Users/sbianchi/Documents/tex/DUST} 

\begin{appendix}
\section{Metallicity gradients}
\label{app:metgrads}

The most reliable determination of gas metallicity is through the so-called direct method, namely, by measuring the physical parameters of the gas: 
the electron temperature from the ratio of a weak auroral line (e.g., [OIII] 4363 \AA) and a strong emission line from the same ionic species 
(e.g., [OIII] 5007 \AA); the electron density, from the ratio of two lines of the same ionic species very close in $\lambda$ (e.g., [SII] 6717, 6731 \AA\,; see, e.g., \citealt{OsterbrockBook1989}).
From  strong lines (e.g., [OIII] and [OII]  for O/H) and  hydrogen recombination lines, solving the equation of the statistical equilibrium equations at the 
provided conditions of temperature and density, we can derive  the ionic abundances. The metallicity is then obtained summing up the ionic abundances, 
eventually correcting for the ionic abundances of the unseen ionization stages. 
After estimating the metallicity in various \ion{H}{ii} regions along the disk of a galaxy, a metallicity gradient can be determined.
Unfortunately, only a few galaxies have gradients obtained with the direct method: in the recent compilation and analysis by \citet{ZuritaMNRAS2021},  O/H gradients are 
provided for 12 galaxies only, only 4 of which are in our sample. Instead, gradients for all our targets are available from \citet{PilyuginAJ2014}. They have been derived from 
measurements of strong lines only, adopting empirical methods that perform well when compared to gradients obtained with the direct method \citep{PilyuginMNRAS2011,PilyuginMNRAS2012}.

From an extensive compilation of literature and archival line emission measurements, \citet{DeVisA&A2019} derived metallicities for a large fraction of DustPedia galaxies. These authors used a variety of strong-line calibration methods and provided characteristic metallicities (at $0.4\times R_{25}$) and metallicity gradients (adopting average gradients as priors for galaxies with few measurements across the disk). It is a well known fact that the estimate of the gradient depends on the adopted metallicity calibrator \citep[see, e.g.,][]{ZuritaMNRAS2021}. For one of our targets, NGC~5194, the gradient from direct measurements  \citep{BergApJ2020,ZuritaMNRAS2021} as well as from the strong-line calibration of 
 \citet{PilyuginAJ2014} is negative, as usually found in many galaxies. Instead, the metallicity distribution is flat using the PG16S calibration \citep{PilyuginMNRAS2016}, the preferred choice of  \citet{DeVisA&A2019}. For the N2 method \citep{PettiniMNRAS2004}, which was chosen by \citet{BianchiA&A2019}, the gradient found by  \citet{DeVisA&A2019} for NGC~5194 becomes positive \citep[as also shown by][]{ZuritaMNRAS2021}.
 
\begin{figure*}
\includegraphics[width=\hsize]{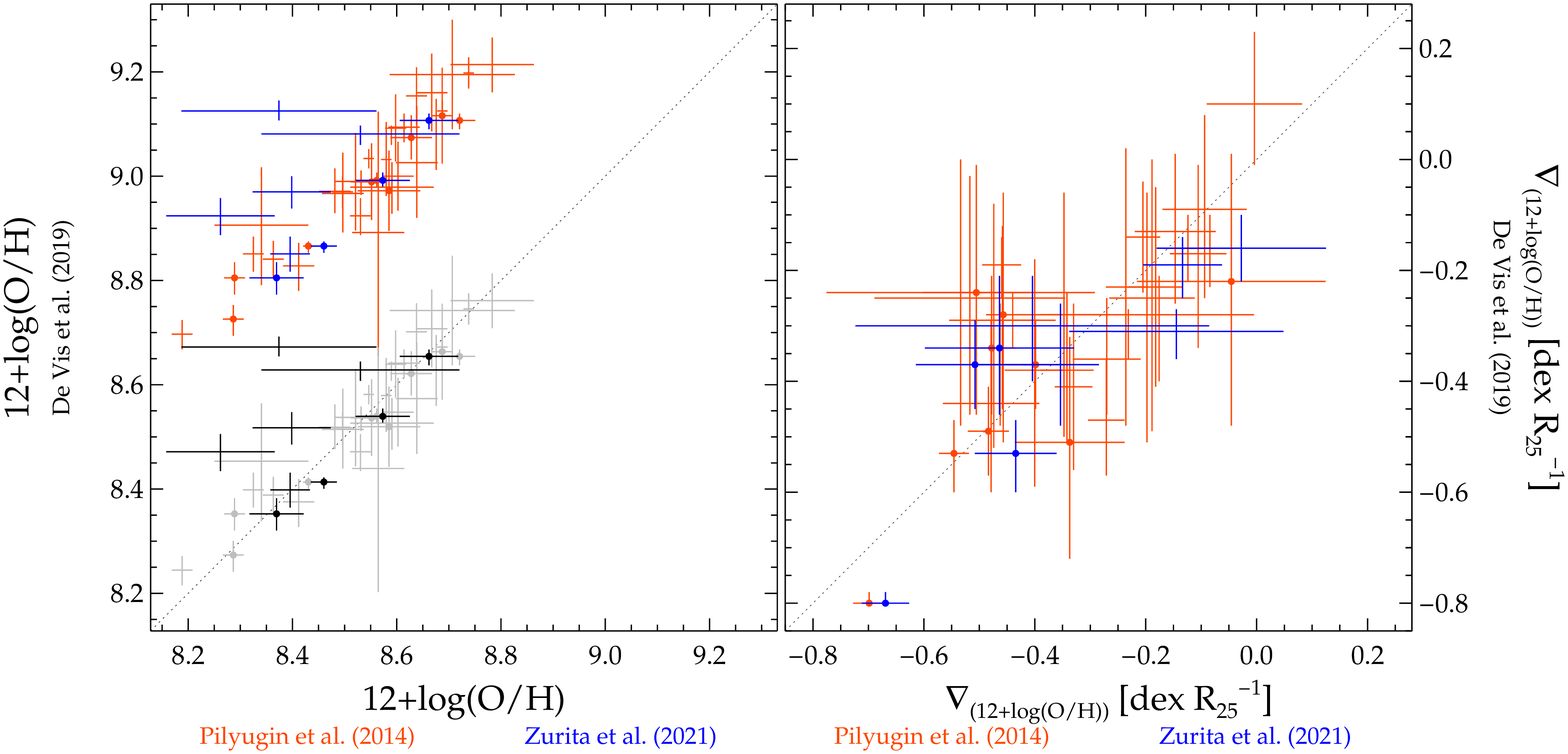}
\caption{
Comparison between the characteristic metallicities (at $0.4\times R_{25}$; left) and metallicity gradients (right) obtained by \citet{DeVisA&A2019} for DustPedia galaxies with the \citet{KobulnickyApJ2004} calibration (on ordinates), and the same values from \citet{PilyuginAJ2014} and \citet[][on abscissae; red and blue datapoints, respectively]{ZuritaMNRAS2021}. For each galaxy, characteristic metallicities and gradients on abscissae have been derived from the original data using the same value for $R_{25}$ \citep[from][]{DeVisA&A2019}. Grey and black datapoints on the left panel show the comparison with  \citet{PilyuginAJ2014}  and \citet{ZuritaMNRAS2021}, respectively, once the characteristic metallicity of \citet{DeVisA&A2019} has been corrected by 0.45 dex.  \citet{DeVisA&A2019} have 58 galaxies in common with \citet{PilyuginAJ2014} \citep[9 with ][]{ZuritaMNRAS2021}. Datapoints for the galaxies used in this work are shown with a circle.
}
\label{fig:compgrads}
\end{figure*}

For the DustPedia galaxies in common between the two samples, we compared the characteristic metallicities and gradients obtained by \citet{DeVisA&A2019} with those from \citet{PilyuginAJ2014}. When using the N2 or PG16S calibrations, the characteristic metallicities from \citet{DeVisA&A2019} are within 0.05 dex of those from \citet{PilyuginAJ2014}. The smallest difference and scatter in the gradients, instead, is for the KK04 theoretical calibration of \citet[][see Fig.~\ref{fig:compgrads}, red datapoints]{KobulnickyApJ2004}. The metallicities obtained with this methods have higher values, but once they are corrected by 0.45 dex, they show a small scatter with respect to those  from \citet{PilyuginAJ2014}. A similar comparison is shown in Fig.~\ref{fig:compgrads} with \citet{ZuritaMNRAS2021}, however, for a smaller number of objects (blue and grey datapoints).
 
\begin{figure}
\includegraphics[scale=0.3]{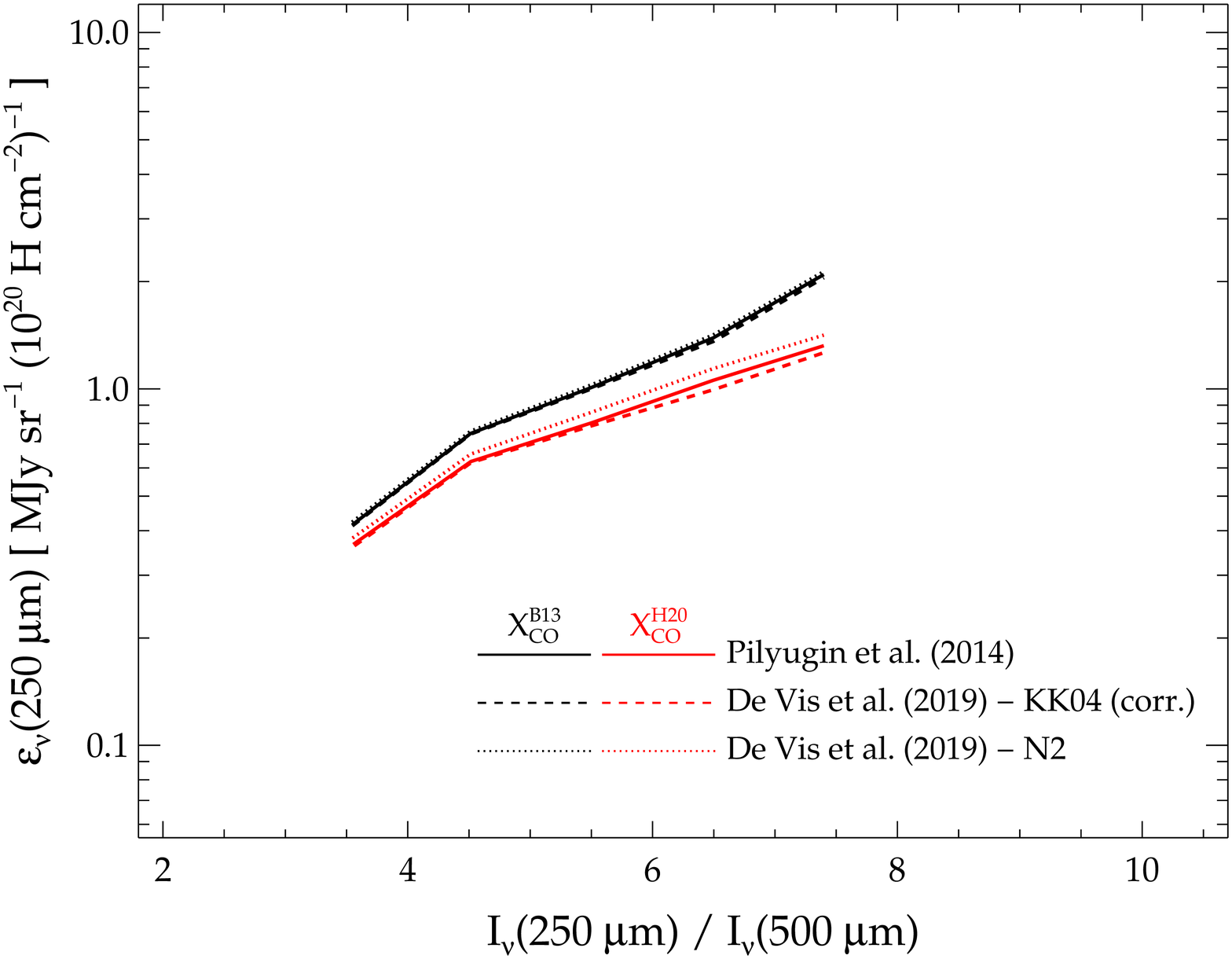}
\caption{
Average $\emy$ versus  $\emyrat$ for different  choices of  $X_\mathrm{CO}$ and  metallicity gradients.
}
\label{fig:diffgrads}
\end{figure}

In this work we used the metallicity gradients from \citet{PilyuginAJ2014}: this is the more straightforward choice, because of the "regular"
gradient for NGC~5194 and because there is no need for corrections as for the KK04 metallicities of \citet{DeVisA&A2019}. 
Furthermore, this choice resulted in a slightly smaller scatter of $\emy$. Nevertheless, the exact choice of the
metallicity gradient appears to have little impact on the results of this work, as shown by the small differences  in the average emissivities in Fig.~\ref{fig:diffgrads}.

\section{Effects of ISRF mixing}
\label{app:bias}

The emissivity is a useful benchmark for models if dust emission comes from grains heated by a uniform ISRF. If instead observations include emission from grains exposed to a distribution of starlight intensities, the spectral shape of the emissivity might be affected by the range in ISRFs; it is then more difficult to extract from the SED the signatures of grain optical properties and size distributions. We explore here the effects of the mixing of ISRF intensities on the FIR-to-submm emissivities, using the THEMIS model and the DustEM code \citep{CompiegneA&A2011} to predict dust emission under different heating conditions.

\begin{figure*}
\centering
\includegraphics[scale=0.3]{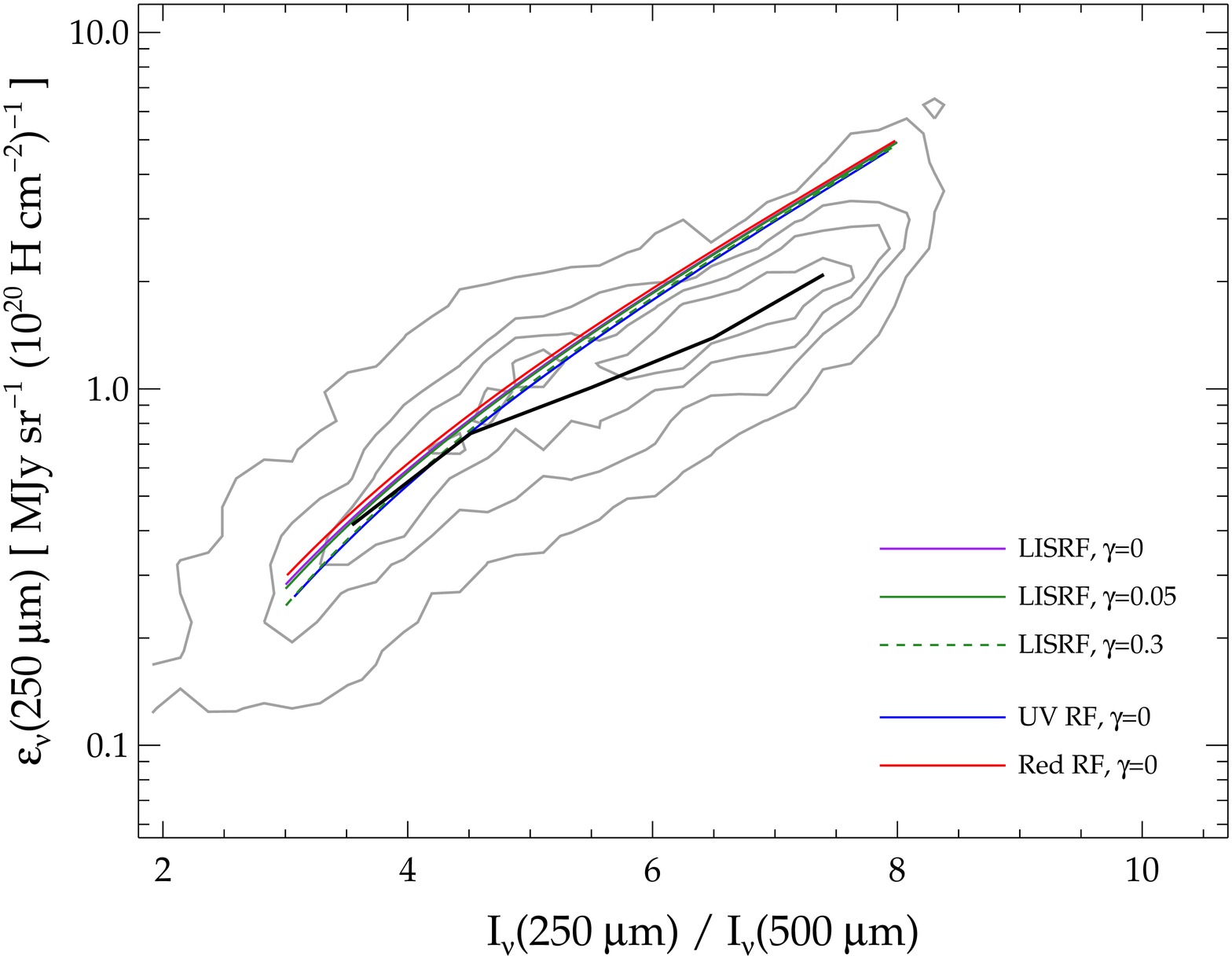}~~\includegraphics[scale=0.3]{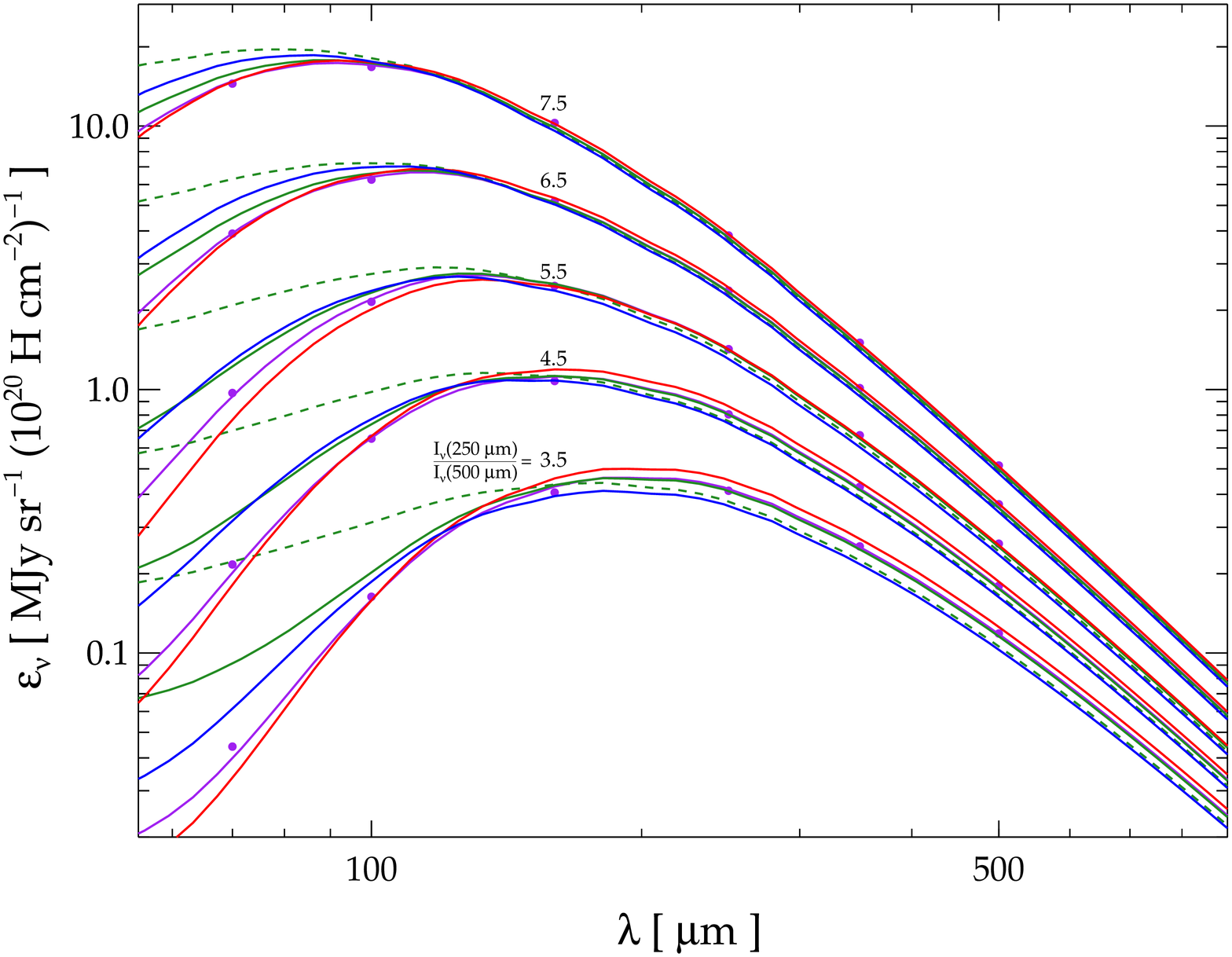}
\caption{
Variations of  $\epsilon_\nu$ for the THEMIS model under various heating conditions.
Left panel: $\emy$ versus $\emyrat$  for radiation fields scaled on the LISRF, with different values
for $\gamma$, the fraction of dust heated by a radiation field distribution ($\alpha=2$ and $U_\mathrm{max}=10^7$ for $\gamma \ne 0$; Eq.~\ref{eq:dl07}); and for 
ISRFs dominated by UV (young stars) or red (old stars) radiation ($\gamma=0$; see text for details).
Contour plots (gray lines) and averages (black line) from the analysis in the main text are shown for reference (Fig.~\ref{fig:emy250_xcob}).
Right panel: $\epsilon_\nu$ versus $\lambda$ for the central values of  the $\emyrat$ bins and the same cases as in the left panel. 
Dots represent the emissivities integrated over the {\em Herschel} filter response functions (for the LISRF, $\gamma=0$ case only).
}
\label{fig:emy250_xco_b13_test}
\label{fig:sed_test}
\end{figure*}

The reference model used in the main text consists of the simple case of the THEMIS dust mixture heated by a uniform ISRF. 
Usually, the radiation field heating the dust is scaled on the local one, the LISRF of \citet{MathisA&A1983}, by setting $\mathrm{ISRF} = U \times \mathrm{LISRF}$. In Fig.~\ref{fig:emy250_xco_b13_test} (left panel), we show $\emy$ versus $\emyrat$ for values of the scale factor ranging from $U=0.1$ to $20$. As $U$ grows the ISRF becomes more intense, and both $\emy$ and $\emyrat$ increase. In Fig.~\ref{fig:sed_test} (right panel) we show the emissivity SEDs for the five $\emyrat$ bins used in the analysis: from the lower to the upper, SEDs have $\emyrat = 3.5, 4.5, 5.5, 6.5,$ and 7.5, corresponding to $U=0.35, 1, 2.8, 8.2,$ and 26.

\citet{DraineApJ2007b} introduced the consuetude of describing the SED of galaxies as the sum of a dust component heated by a radiation field of intensity $U_\mathrm{min}$ and another, hotter, component heated by a power-law distribution with $U_\mathrm{min} < U < U_\mathrm{max}$ (as proposed by \citealt{DaleApJ2001}). It is
\begin{equation}
\frac{1}{M_\mathrm{d}} \frac{dM_\mathrm{d}}{dU} = (1-\gamma)\delta(U-U_\mathrm{min})+\gamma \frac{(\alpha-1)U^{-\alpha}}{U_\mathrm{min}^{1-\alpha}-U_\mathrm{max}^{1-\alpha}},
\label{eq:dl07}
\end{equation}
where $\gamma$ is the fraction of dust heated by the ISRF powerlaw distribution and $\alpha$ its spectral index (Eq.~\ref{eq:dl07} being formally correct for $\alpha>1$). When fitting global SEDs, $M_\mathrm{d}$ is the total dust mass; for resolved analyses, it is the dust mass within the resolution area. Typically, $\alpha\approx 2$ both in fits of global \citep{DaleApJ2012} and resolved SEDs \citep{AnianoApJ2020}. We follow \citet{AnianoApJ2020} and use $U_\mathrm{max}=10^7$, although for $\alpha=2$ and the wavelengths of interest here there is little difference if $U_\mathrm{max}$ is set as low as $10^4$.

The case of a uniform ISRF is equivalent to adopting $\gamma=0$ in Eq.~\ref{eq:dl07}. When global SEDs are analyzed, $\gamma$ is found to be a few percent and the dust heating 
(and emission) is still dominated by the $U_\mathrm{min}$ component \citep{DaleApJ2012}. The composite emissivity derived with 
$\gamma=0.05$ is shown in Fig.~\ref{fig:emy250_xco_b13_test}. The $\emy$ versus $\emyrat$ trend is almost unaltered with respect to the reference $\gamma=0$ case, although the average radiation field corresponding to each $\emyrat$ bin becomes larger, $\langle U\rangle=0.53, 1.5, 4.1, 12,$ and 36 (see the aforementioned references for the mathematical definition of $\langle U\rangle$). Emissivity values for $\lambda \ge 160 \, \mu$m are within less than 2\% of those for a uniform ISRF, while values at the shorter {\em Herschel} wavelengths could be much larger, though the difference reduces for the more intense ISRFs (and larger emissivities).  

The contribution of the power-law distribution, instead, is found to be larger in resolved studies: for the galaxies in common with our sample, and at a comparable resolution to the one adopted here, it is $\langle \gamma \rangle \approx 0.3$ in the work of \citet{AnianoApJ2020}.  Again, the $\emy$ versus $\emyrat$ trend changes little, although the reference $\emyrat$ values are now produced by $\langle U\rangle=1.1, 3.3, 9, 26,$ and 79. The difference with the uniform ISRF heating is now larger: the emissivity in the three SPIRE bands is $\approx 10\%$ smaller for  $\emyrat=3.5$ and 7\% for 4.5, reducing to less than 5\% for larger $\emyrat$; at shorter wavelengths, instead, the emissivity is larger (up to a factor 2 at 160 $\mu$m for $\emyrat=3.5$). For a given $\emyrat$, the general effect of the mixing of radiation fields is then that of  a broadening of the SED, with relatively small changes in the submm emissivity,
but larger changes for $\lambda \la160 \mu$m and a shift of the peak emission towards smaller wavelengths.

The basic assumption in the formalism we use here is that the spectral shape of the ISRF does not change, as it is always the same as that of the LISFR. \citet{DraineApJ2021} investigated the effects of different spectra and found that emission in the mid-infrared can change significantly if the ISRF is richer in UV radiation. However, at the wavelengths 
of interest here, the spectral shape of the radiation field has smaller effects: this is shown in Fig.~\ref{fig:emy250_xco_b13_test}, where we plot the cases for the
two extreme spectra in the work of \citet{DraineApJ2021},  a UV-rich 3~Myr old starburst, and a UV-poor, red, spectrum from a very old stellar population (used by \citealt{GrovesMNRAS2012} to describe the bulge population of M~31). Changes in the emissivity (which are stronger for the UV-rich heating radiation field) are still within those
we have considered so far, in particular for the larger $\emyrat$ where observations diverge from our reference model. We do not consider this case any further, and revert to the standard scaling on the LISRF.

\begin{figure*}
\centering
\includegraphics[scale=0.3]{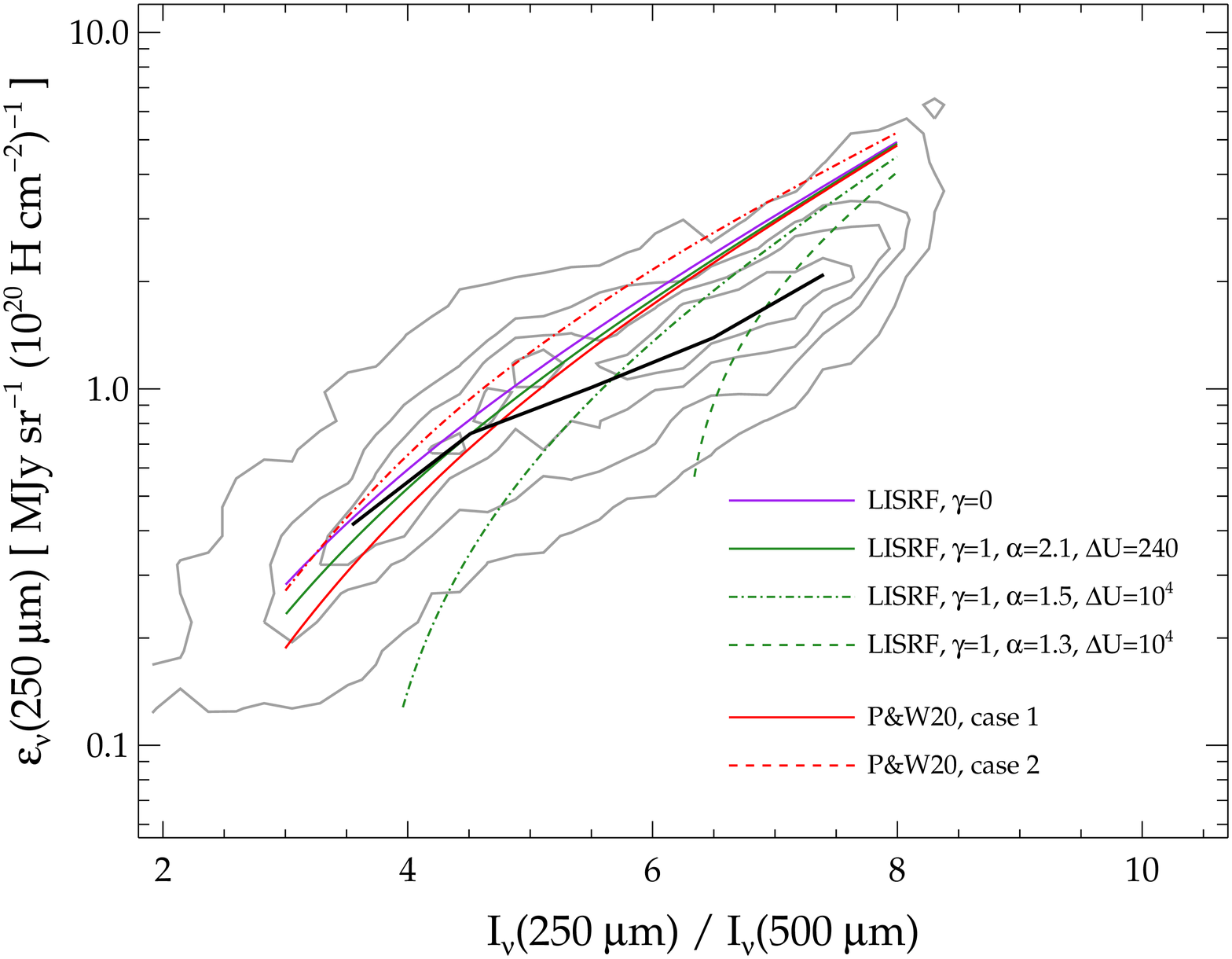}~~\includegraphics[scale=0.3]{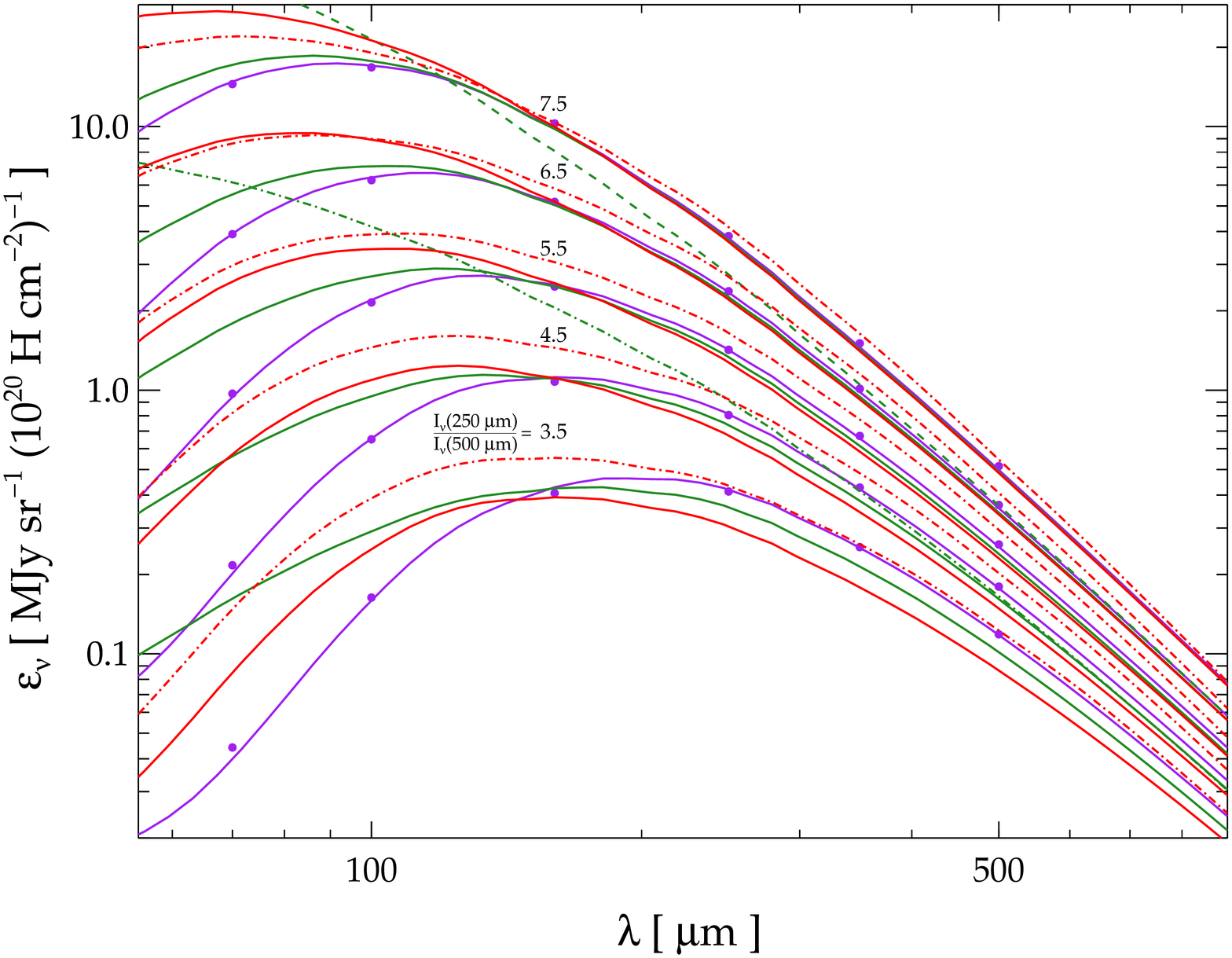}
\caption{
Same as Fig.~\ref{fig:emy250_xco_b13_test}, but for different radiation mixes. Beside the reference $\gamma=0$ case, we show results for  $\gamma=1$ and
various values for $\alpha$ and $\Delta U$; and for models taking into account the  biases described by \citet{PriestleyMNRAS2020}, assuming that grains exposed to the 
attenuated radiation field have the THEMIS properties (case 1) or those of the CMM grain mixture (case 2;  see text for details).
For clarity, in the left panel the $\alpha=1.5$ SED is shown for $\emyrat=5.5$ only, and the $\alpha=1.3$ SED for $\emyrat=7.5$ only. 
}
\label{fig:emy250_xco_b13_test2}
\label{fig:sed_test2}
\end{figure*}

An alternative ISRF distribution is used by \citet{GallianoA&A2021}, neglecting the $\delta(U-U_\mathrm{min})$ component of Eq.~\ref{eq:dl07} (or, equivalently, setting $\gamma=1$).
The shape of the dust SED illuminated by this distribution can then vary as a result of changes in $U_\mathrm{min}$, $\alpha$, and in the U range, via the parameter $\Delta U =
U_\mathrm{max}-U_\mathrm{min}$. By fitting the global SEDs of 798 DustPedia galaxies, \citet{GallianoA&A2021}  find median values $\alpha\approx 2.1$ and  $\Delta U \approx 240$.
The emissivities for this case are shown in Fig.~\ref{fig:emy250_xco_b13_test2}. In the SPIRE bands, $\epsilon_\nu$ is biased low by $\approx$15\% and 10\%
for $\emyrat=3.5$ and 4.5, respectively, while the $\emyrat$ bins now corresponds to ISRF intensities $\langle U\rangle=0.5, 1.5, 4.1, 11$ and 32. Again, the trends we obtain for this
model are not too different from those of the reference one and do not reproduce the observed averages at larger $\emyrat$ values. While waiting for an analysis of resolved SEDs
using the \citet{GallianoA&A2021} formalism, we cannot exclude the presence of broader ISRF distributions in localized regions of a galaxy. In particular, by choosing  $\alpha < 2$
and appropriate $\Delta U$ values, we can have $\emy$ versus $\emyrat$ values approaching the observations: we show in the left panel of Fig.~\ref{fig:emy250_xco_b13_test2} two cases
with $\Delta U=10^4$, one for $\alpha=1.5$, passing close to the observed average at $\emyrat=5.5$ and one for $\alpha=1.3$, near the $\emyrat=7.5$ value (for a given $\alpha$,
the low-$\emy$ part of the trend shifts to the right as $\Delta U$ increases). Despite matching $\emy$, however, these two cases cannot reproduce the emissivity SED, since they have emission peaks at wavelengths much shorter than what we find (Fig.~\ref{fig:emy250_xco_b13_test2}, right panel).

The power-law distribution in Eq.~\ref{eq:dl07} was proposed by \citet{DaleApJ2001} as a way to describe different
heating conditions, with $\alpha$ ranging from 2.5 for a diffuse, uniform medium where the ISRF intensity decreases 
mainly with the increase of the distance between dust and the sources and up to $\alpha=1$ for dense clouds where the ISRF intensity
reduces because of dust absorption. While a proper radiative transfer model is needed to investigate the changes
in the heating fields due to dust attenuation (see, e.g.,\ \citealt{YsardA&A2012} for the variations in the dust emissivity of externally illuminated clouds
of size up to a few parsecs), a simpler scenario has been proposed by \citet{PriestleyMNRAS2020}:
they suggest that, along lines of sight of high ISM column density, a considerable fraction of the dust emission 
comes from grains heated to colder temperatures by a radiation field attenuated by dust itself.
Using their formalism, we explored the case for column density $\Sigma_\mathrm{ISM}=$100 M$_\odot$ $\mathrm{pc}^{-2}$, 
corresponding to a dust optical depth in the V-band $\tau_\mathrm{V} \approx 6$ for the average MW properties \citep{BohlinApJ1978}.
We adopted the rather extreme case in which 25\% of the dust emission is due to dust heated by an unattenuated ISRF of intensity $U$, and 75\% to dust heated by an attenuated 
radiation field which, assuming the MW extinction law and that all emission is behind a screen with $\tau_\mathrm{V}=6$,  has a reduced global intensity corresponding to about
$0.15\times U$ (though with a redder spectral shape with respect to the LISRF). We show in Figure \ref{fig:sed_test2}, the results using the THEMIS model for dust exposed to the 
unattenuated field, and two different properties for dust at higher density: in the first case, we  again use THEMIS; in the second case, we use the CMM model of 
\citet{KoehlerA&A2015}, which takes into account dust growth and coagulation in denser media. When using the THEMIS model for the higher density environment, the estimate of 
the emissivity is $\approx$20\% lower in the SPIRE bands for $\emyrat = 4.5$; for CMM dust,  the mixing of heating fields conjures with the higher grain cross sections, and the difference is smaller ($\epsilon_\nu$ is higher by 15\%). In any case, the
 trend of $\emy$ versus $\emyrat$ is not significantly altered from that of the single $U$ case; the bias for $\lambda \ge 160\, \mu$m and $\emyrat > 5$ is $\approx 15\%$ at maximum.
\citet{KoehlerA&A2015} also present other dust aggregate models which, however, result in higher emissivities with respect to CMM (and THEMIS; see also
 \citealt{BianchiA&A2019}) and thus cannot be reconciled with the observed mean $\emy$ at $\emyrat=4.5$ and the trend for $\emyrat>5$.

In conclusion, the mixing of ISRF intensities is not likely to bias strongly the estimate of $\epsilon_\nu$ for $\lambda\ge 160\mu$m, at least for $\emyrat > 5$. 
It would have been possible to  use $\epsilon_\nu(160 \,\mu\mathrm{m})$ as a reference value, as it shows the lower bias,  $\le 6\%$ in most test cases shown here.
We prefer nevertheless $\emy$ 
since it is a common reference in the literature and it has been used for the global emissivity estimates of \citet{BianchiA&A2019} and derived directly  in the foreground of the Virgo Cluster  from SPIRE observations of the MW cirrus (which was undetected in PACS images; \citealt{BianchiA&A2017}).

\end{appendix}

\end{document}